\documentclass[prb,amsmath,amssymb,reprint]{revtex4-1}
\usepackage[]{graphicx}
\usepackage[]{amsmath}
\usepackage[]{bm}
\usepackage[]{microtype}
\usepackage[]{siunitx}
\usepackage[]{xcolor}
\usepackage{footnote}

\begin{document}
\title[]{Atomistic Structure Learning Algorithm with surrogate energy model relaxation}
\author{Henrik Lund Mortensen}
\author{S{\o}ren Ager Meldgaard}
\author{Malthe Kj{\ae}r Bisbo}
\author{Mads-Peter V.\ Christiansen}
\author{Bj{\o}rk Hammer}
\email{hammer@phys.au.dk}
\affiliation{ 
Department of Physics and Astronomy, Aarhus University, DK-8000 Aarhus C, Denmark.
}
\date{\today}

\begin{abstract}
  The recently proposed Atomistic Structure Learning Algorithm (ASLA)
  builds on neural network enabled image recognition and reinforcement
  learning. It enables fully autonomous structure determination when
  used in combination with a first-principles total energy calculator,
  e.g.\ a density functional theory (DFT) program. To save on the
  computational requirements, ASLA utilizes the DFT program in a
  single-point mode, i.e.\ without allowing for relaxation of the
  structural candidates according to the force information at the DFT
  level. In this work, we augment ASLA to establish a surrogate energy
  model concurrently with its structure search. This enables
  approximative but computationally cheap relaxation of the structural
  candidates before the single-point energy evaluation with the
  computationally expensive DFT program. We demonstrate a
  significantly increased performance of ASLA for building benzene
  while utilizing a surrogate energy landscape. Further we apply this
  model-enhanced ASLA in a thorough investigation of the
  $c(4\times 8)$ phase of the Ag(111) surface oxide. ASLA
  successfully identifies a surface reconstruction which has
  previously only been guessed on the basis of scanning tunnelling
  microscopy images.
\end{abstract}
\maketitle
\section{Introduction}

Structure determination is a cornerstone in the development of new
functional materials for applications as photo-voltaics, thermoelectrics,
electrolytes, and
heterogeneous catalysts \cite{materialrs}. Often, advances in creating new such
materials are lead by experimental discoveries and subsequent
characterization of the compounds serve to identify their precise
structure and composition at the atomic scale. Once the
structure of a material is known, its physico-chemical properties may
be rationalized on theoretical grounds providing a basis for further
improvements. The process of structure determination is highly
non-trivial and is hampered by the rapid growth with problem size of
both the configurational space and the cost of the individual first
principles quantum mechanical computer calculations required.

A number of strategies for automated structural search have
emerged. These include simulated annealing based on molecular dynamics
(MD) \cite{sim_an} and stochastic search based on e.g.\ basin hopping \cite{basin_hop}
or even evolutionary algorithms \cite{GA1,GA2,GA3,GA4,GA5,GA6}. The recent developments in
machine learning (ML) techniques in chemical physics \cite{Schmidt_review_2019} have
successfully been introduced to lower the computational demands for
these search strategies. One method has been to model the potential
energy landscape accurately with kernel based methods \cite{K1,K2,K3} or neural
networks \cite{NN1,NN2,NN3,NN4} so that many expensive energy and force
evaluations necessary for the MD and the local relaxations can be
circumvented. Typically, the machine learning models are trained
on-the-fly with active learning protocols
\cite{SS:alexandrova,active:rinke,
  active:oguchi,active:ove,GAPRSS:boron,GAPRSS:crystal,activeSS:calypso,activeSS:shapeev,SS_dftb:maxime,activeFF:roitberg,activeFF:Zhang,LEA,activeMD:Vita,activeMD:andrew,activeMD:Ohno,activeMD:evgeny,activeMD:kresse,activeSS:deringer,NEB_hannes2017,NEB_hannes2017,localOpt_karsten2019}
and improved as more
and more first-principles datapoints are accumulated. Other means for
speeding up structural searches with machine learning have been the
introduction of acquisition functions, based on Bayesian statistics,
balancing exploration and exploitation \cite{explo1,explo2}, the clustering analysis
in the selection of evolutionary populations \cite{cluster}, the estimation of
local energies to guide mutation and cross-over operations in
evolutionary searches \cite{le1,le2,le3}, and the construction of
artificially convex energy landscapes for initial relaxation of
evolutionary candidates \cite{convex1,convex2}.

Notwithstanding the considerable computational speedups that ML
techniques have lead to in structural determination, the underlying
search paradigm remains of stochastic nature. This means that as a
catalogue of more and more likely structural candidates is
constructed, no knowledge or understanding of the bonding mechanism is
developed and identifying more candidates still relies on the element
of chance. To tackle this issue, we have recently proposed the
Atomistic Structure Learning Algorithm (ASLA) \cite{asla}, which uses
neural network based image recognition and reinforcement learning to
iteratively construct the most stable atomic structure given only the
cell and the stoichiometry. ASLA is designed to develop a rational
search strategy in which interatomic arrangements are introduced
because they appear plausible based on prior data. ASLA reads
discretized 2D or 3D image representations of molecular compounds and
solids. It builds the next promising structure, conducts a
single-point, i.e.\ an unrelaxed, first-principles energy calculation,
and uses this new structure-stability datapoint to train the neural
network.

In this work, we extend ASLA to include a surrogate energy
landscape that is constructed on-the-fly. Whenever ASLA proposes a new
structure, the structure is subjected to a computationally inexpensive
local relaxation in the model before the computationally
expensive first-principles single-point calculation is conducted. The
method is shown to dramatically increase the performance of ASLA when
applied to solving for the most stable C$_6$H$_6$-isomer, which is
benzene.  We further apply ASLA with surrogate model relaxations to
the problem of finding the most stable geometry and stoichiometry of
the Ag(111) $c(4\times 8)$-phase surface, which has been experimentally and
theoretically investigated in Refs.\ \onlinecite{schnadt_prb_2009,martin_j_phys_chem_c_2014}. Ag(111) oxide surfaces are known
to be catalysts for the important ethene epoxidation reaction and
have been studied for decades. Yet the structure of such systems
continue to surprise \cite{ag111_reuter}. The important catalytic properties
together with the rich structural diversity calls for reliable,
automated global search algorithms for structure determination.

The paper is outlined as follows.
In the first section, we introduce the details of the model. In the
second section, we demonstrate the performance increase for a simple
benzene system, where many restarts can be made and good statistics obtained.
Lastly, in the the final section, we employ our algorithm to identify the
aforementioned Ag(111) oxide surface structure. Doing so, we employ the
method for a large number of different silver oxide stoichiometries.
The most stable structure is identified and discussed.

\section{ASLA}

\begin{figure}
  \centering
  \includegraphics{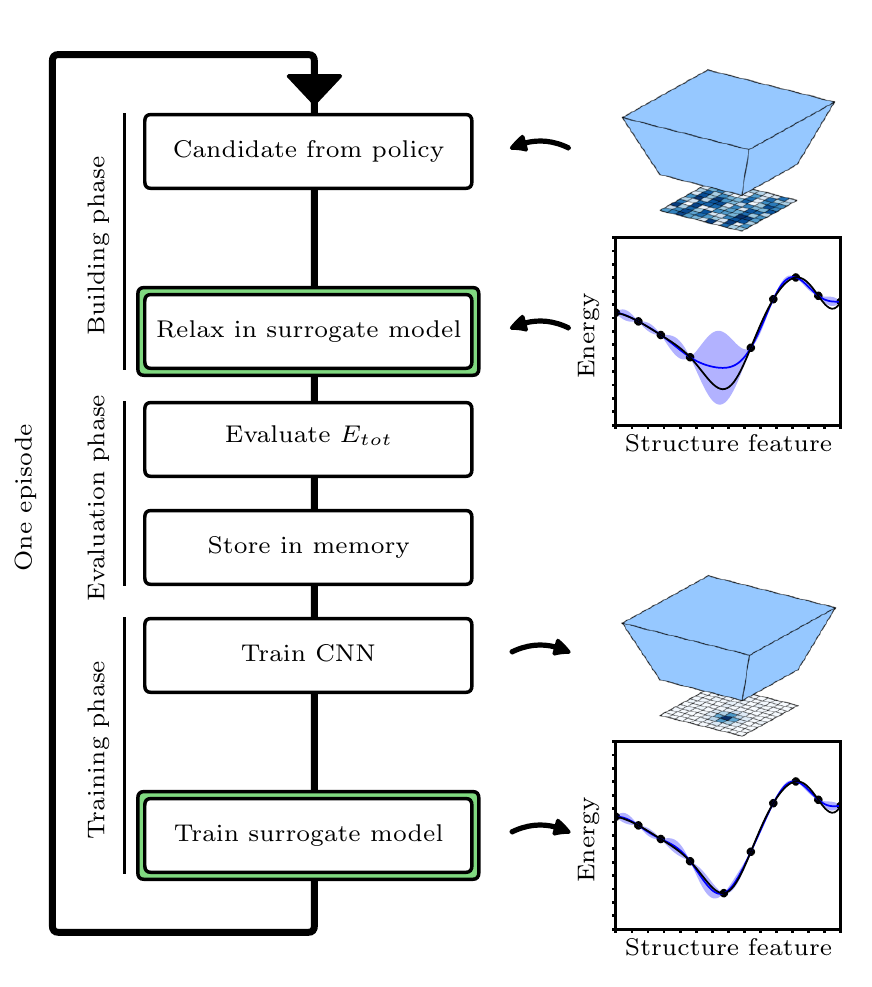}
  \caption{Flowchart showing the three phases of ASLA. The green nodes
    highlight the contributions from this paper. The blue polygon
    represents the CNN producing the $Q$-value map, while the
    surrogate energy model, in which the relaxation is performed, is
    schematically represented as the blue curve in the
    structure-energy plot. The true energy curve is shown in black and
    black dots are training points, while the shaded area marks the
    model uncertainty. Both the CNN and the surrogate energy model
    improve in each episode, leaving ASLA better equipped for building
    low-energy structures in later episodes. }
  \label{fig:asla_flow}
\end{figure}

In the Atomistic Structure Learning Algorithm (ASLA) \cite{asla,asla_pseudo_3D,asla_representation}, the global
optimization problem of structure search is formulated as a
reinforcement learning problem, where an \emph{agent} learns 
to build the most stable atomistic structure. The learning
appears in \emph{episodes}, in each of which, the agent itself creates a new
prospective atomistic structure whose total energy is
provided by an external density functional theory (DFT)
program \cite{gpaw1,gpaw2}. Each learning episode contains
three distinct phases, the \emph{building phase}, the \emph{evaluation phase},
and the \emph{training phase} as illustrated in the
flowchart given in Fig.\ \ref{fig:asla_flow}.

\begin{figure*}
  \centering
  \includegraphics[width=0.9\textwidth]{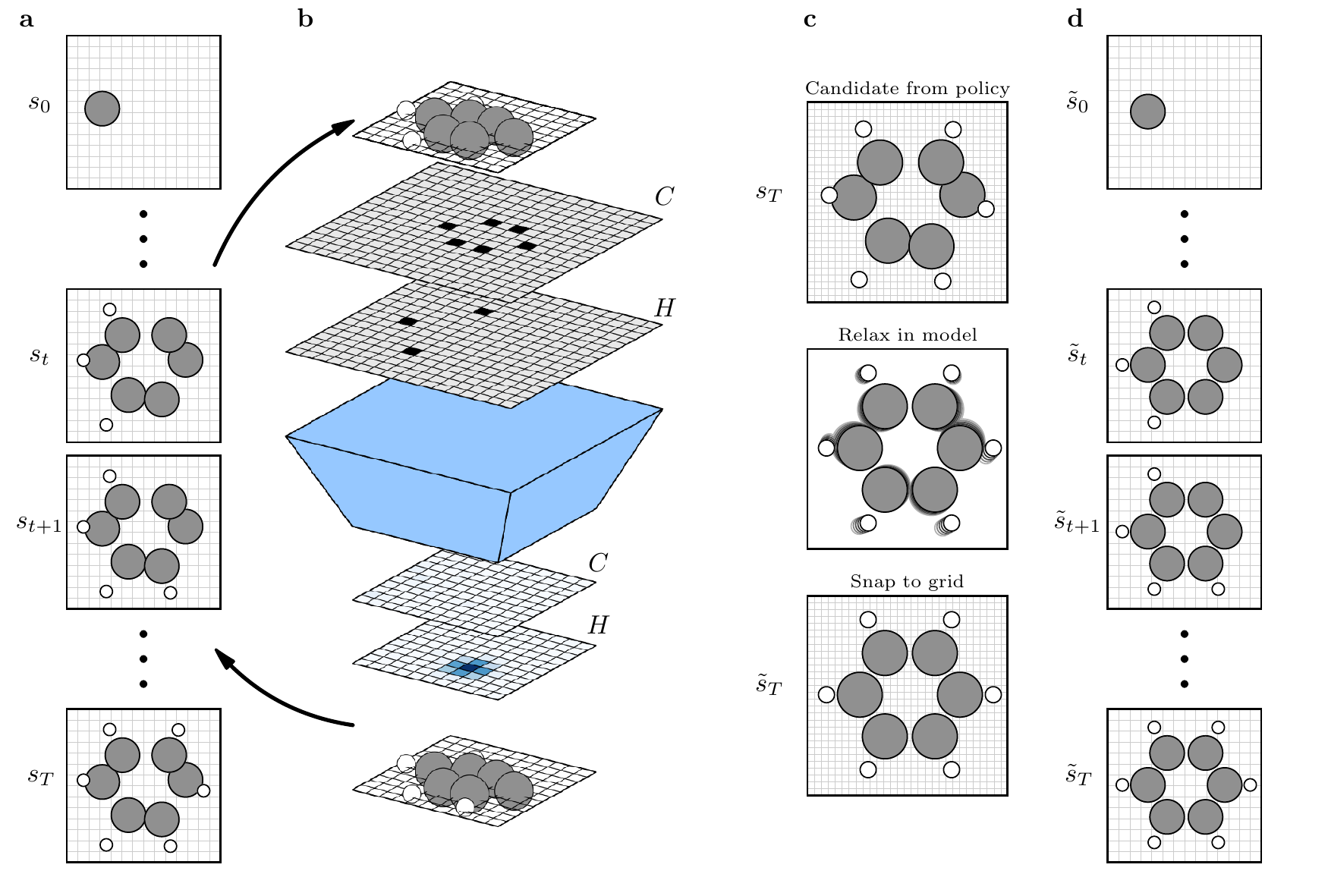}  
  \caption{\textbf{a} A candidate structure is constructed
    atom-by-atom in the \emph{candidate-from-policy} operation by iteratively
    consulting a CNN (shown in \textbf{b}) and following the modified
    epsilon-greedy policy described in the text. This results in the
    sequence of structures, $s_0,\dots,s_T$. \textbf{c} The final
    structure, $s_T$, is relaxed in the surrogate energy model,
    resulting in the $\tilde s_T$-structure, for which the total
    energy is evaluated. \textbf{d} The $\tilde s_T$-structure is
    decomposed atom-by-atom to obtain the sequence,
    $\tilde{s}_0,\dots,\tilde{s}_T$, which is stored in the ASLA
    memory together with the energy, and used for training of both the
    CNN and the surrogate energy model. }
  \label{fig:build_and_relax}
\end{figure*}

In the building phase, ASLA commences by constructing a new structural
candidate by following the policy as guided by the
current state of the agent. 
It starts from a \emph{template} structure, $s_0$, (possibly an empty
computational cell) and proceeds iteratively
to a final state, $s_T$, such that a total of $N_\text{atom}$ atoms
have been placed, as further detailed in Fig.\
\ref{fig:build_and_relax} \textbf{a}.
In each iteration the intermediate state, $s_t$, is given
to the agent equipped with a convolutional neural network (CNN), that
responds with its expectation, $Q(s_t,a_t)$, of the reward for
the most stable final structure attainable if \emph{action}, $a_t$, is
taken. An action is the combined
information about position and type of the next atom to be placed,
and a reward is a function whose maximum coincides with
the most stable final structure found at any time, see
Refs.\ \cite{asla,asla_pseudo_3D} for details. With discretization of space,
all possible actions for each type of atom can be represented on a 2D or 3D
grid depending on the problem being solved. The expected rewards for all possible
actions can hence be represented as $Q$\emph{-value maps} on such grids.
Discretized atomistic structures and $Q$-value maps thus
share the same data structure as illustrated in Fig.\
\ref{fig:build_and_relax} \textbf{b}. 
The $Q$-value maps form together with a \emph{policy} the basis for a
decision process regarding the actual action, $a_t$, taken at state
$s_t$ leading to state $s_{t+1}$.  We employ a modified epsilon-greedy policy. This means that
ordinarily it is the expected most rewarding action (greedy), which is chosen,
but occasionally with some small likelihood (epsilon) a stochastic
element is used in choosing the action \cite{asla}.

After constructing a candidate from the policy we add in the present
version of ASLA a new element, namely a \emph{structural
  relaxation}. The changes that a structural candidate undergoes
during this are illustrated in Fig.\ \ref{fig:build_and_relax}
\textbf{c}.  The structure relaxation seeks to identify the nearest
local minimum energy structure in the vicinity of the structure just
built. To avoid any significant computational expense of this
operation, it is carried out in a surrogate energy landscape described
in the next section. After completing the relaxation, the atom
positions are adjusted slightly (snapped to grid) to ensure
that the relaxed structure, now $\tilde{s}_T$, can be represented on
the grid used for training the agent.

The reinforcement learning episode proceeds with the
evaluation phase which is a single-point DFT calculation of the
total energy, $E_\mathrm{tot}$, of structure, $\tilde{s}_T$.  This is
the presumed computationally most intensive operation of the learning
episode. Once finished, a set of state-action-energy datapoints for each iteration of
the candidate-from-policy construction of this episode may be stored in the
\emph{memory} in the \emph{store-in-memory event}. Since the candidate construction
was proceeded by some relaxation,
the state-action pairs will most likely deviate from the
actual states, $s_t$, visited and actual actions, $a_t$, taken during the build
action. Assuming, however, that atoms are placed in the same order, the
state-action pairs, $(\tilde{s}_t, \tilde{a}_t)$, that would have led directly
to the snapped, relaxed structure, $\tilde{s}_T$, can be determined,
cf.\ Fig.\ \ref{fig:build_and_relax} \textbf{d},
and stored with $E_\mathrm{tot}$ in the memory.

The reinforcement learning episode continues with a training phase,
where the CNN and surrogate energy model are updated. The CNN is
trained by first extracting a batch of state-action-energy pairs from
the most recent, most-favorable and random episodes from the memory,
such that the batch comprises $5\times N_\text{atom}$
state-action-energy pairs. The energies are transformed to reward
values, $r \in [-1,1]$, and the batch is expanded by rotated and
mirrored versions of the state-action-energy pairs. The weights of the
CNN are updated by a backpropagation step with a learning rate of
$10^{-3}$ that decreases the total mean squared error between
the predicted value and the reward for state-action-energy pairs
in the batch.

In the present work, we employ the same architecture for the neural
network as in Refs.\ \cite{asla,asla_pseudo_3D,asla_representation}. It is thus a
convolutional neural network (CNN) with three hidden layers and 10
kernels per layer, all with the leaky-ReLU activation function, except
for the output, where a hyperbolic tangent function is applied. The
structure is represented as a one-hot encoded matrix, which represents
the $xy$- or $xyz$-coordinates in the first two or three dimensions,
and the atom type in a final dimension. Penalizing terms are employed
following Refs.\ \cite{asla,asla_pseudo_3D}.

As a final operation during the learning phase of a reinforcement
learning episode, an improved surrogate energy model is prepared for
the next episode. This is done in the \emph{train-surrogate-model} operation,
which represents the final new element introduced to ASLA in this
work.  The surrogate energy model extracts from the memory
$(\tilde{s}_T, E_\mathrm{tot})$ datapoints and constructs a
structure-energy model. The details of the model are given in the next
section.

Upon starting the reinforcement learning cycle, the neural network representing
the agent can be initialized randomly meaning that ASLA learns completely
autonomously from its interaction with the DFT program with no other input than
the template, $s_0$, and the stoichiometry of the final structure, $s_T$. It
may, however, benefit largely in a transfer learning setting, where network
weights are inherited from a prior ASLA run solving a simpler problem.
Initially, before sufficient structure-energy datapoints have been collected and
stored in memory, a sufficiently reliable surrogate model cannot be
established, and the relax operation is simply skipped.

\section{Surrogate model}

For the surrogate model, a Gaussian process (GP) regression model
\cite{GPs_for_ML} is used.  The model is trained on feature
representations of the final structures and their total
energies. The use of a feature representation means
that the GP model incorporates e.g.\ rotational and translational
energy invariances directly, unlike the CNN model used for the agent,
which has to learn that via the applied data augmentation. The GP
regression model, thus learns the underlying energy landscape more
efficiently from the training data compared to what would be expected
from using a CNN model. However, contrary to using a CNN for the
$Q$-value map evaluation, using a GP model would be computationally
expensive owing to the cost of evaluating the feature representation at every pixel in the map.
It can therefore not form the basis for the policy guided building
step in ASLA,
where a CNN with its inexpensive image representation
is preferred. A further reason why a CNN is preferred
for the $Q$-value map prediction is that $Q$-values depend on the
final structures that can be built from incomplete ones. Thus the
$Q$-value evaluation must predict the consequences of future actions,
something which is better done with the deep learning capabilities of
the CNN.

A GP is specified by its prior mean, $\mu(\bm{x})$, and covariance
function (the kernel), $k(\bm{x},\bm{x'})$, which expresses the
beliefs about the system prior to any observations. The training
consists of inferring the posterior distribution, which takes into account the observed training data, $\mathcal{D} = \{X,\bm{E}\}$. The posterior can be
analytically expressed due to the well-behaved marginalization
properties of the GP. Namely, the posterior is also a GP with
mean function given by
\begin{align}
  \label{prediction}
  E(\bm{x}_*) &= \bm{k}_*^T
                                                 (K+\sigma_n^2\mathbb{I})^{-1}[\bm{E}-\mu(\bm{x})]+\mu(\bm{x}).
\end{align}
where $K=k(X,X)$ is the kernel matrix, $\bm{k}_* = k(X,\bm{x}_*)$
and $\sigma_n$ is a regularization hyperparameter that acts to prevent overfitting
(we use $\sigma_n=3.2\cdot10^{-3}$). The posterior mean is taken as the model
prediction of the energy. Computationally, the training is dominated
by the inversion of the kernel matrix, which scales with the number of data
points as $\mathcal{O}(n^3)$.

To take advantage of the rotational, translational and permutation symmetries
of quantum chemistry, structures are represented by feature vectors that exhibit
these symmetries. In this work, we choose the fingerprint descriptors of Oganov
and Valle\cite{oganov_valle_2009}. 

The kernel is chosen to be a sum of two Gaussians with different length scales, similar to that proposed in \cite{bisbo_hammer_2020}
\begin{align}
  \label{kernel GP}
\hspace{-5pt}  k(\bm{x},\bm{x}') = \theta_0 \big( (1-\beta) e^{-|\bm{x}- \bm{x}'|^2/2\lambda_1^2 } +  \beta e^{-|\bm{x}- \bm{x}'|^2/2\lambda_2^2 } \big)
\end{align}
where $\bm{x}$ are the feature vectors and $\theta_0$ is the maximal
covariance. Using two length scales, $\lambda_1$ and $\lambda_2$, improves the
model by allowing the kernel to capture trends in a large, sparsely sampled
configuration space, while maintaining resolution in smaller, more densely
sampled regions. The prior mean is chosen to be a short-ranged
repulsive potential, which is generally present between atoms. This naturally prohibit atoms to get too close during structure relaxations with the model, which may cause
convergence problems when the total energy is evaluated. The feature
vector is analytically differentiable with respect to the Cartesian
coordinates, which allows us to obtain the predicted force according
to the model. The relaxation is performed by iteratively moving the
atoms based on the predicted force.

In training the model, the hyperparameters,
$(\theta_0,\lambda_1,\lambda_2)$, are optimized by maximizing the log
marginal likelihood \cite{GP:Rasmussen}. For the training data, only the
500 lowest energy structures from the ASLA memory is used. 
This ensures that the model resolves the low-energy part of the
configuration space, while keeping the computational training time small.
No effort is thus made to capture high-energy
structures in irrelevant parts of the configuration space.

The model is not meant to reproduce total energies with high
accuracy. Rather, the purpose of the model is to guide ASLA into local
minima of the configuration space in order to aid the global
search. Therefore, we measure the models performance by the improvement in the number of
ASLA episodes needed to solve a global optimization problem, instead
of an error with respect to a test set.

\section{Benzene example}
Having introduced to ASLA that candidates built from the policy may be relaxed
in a surrogate energy landscape, we now turn to probe how it affects the
overall performance of ASLA. In order to be able to make convincing statistics
we choose to have ASLA solve the problem of building the most stable
molecule given six carbon atoms and six hydrogen atoms, i.e.\ C$_6$H$_6$.
The total energy expression is provided by a density functional based tight binding (DFTB) energy expression using
the DFTB+ implementation \cite{dftb_plus}, since this provides a sufficiently accurate description
that the correct chemical bonds are formed, yet at a much reduced computational
cost compared to a DFT energy expression.
As the solution turns out to be the planar molecule \emph{benzene}, we address
this problem with a 2D space. For this a grid spacing of 0.25 {\AA} is used.

\begin{figure}
  \includegraphics[width=0.45\textwidth]{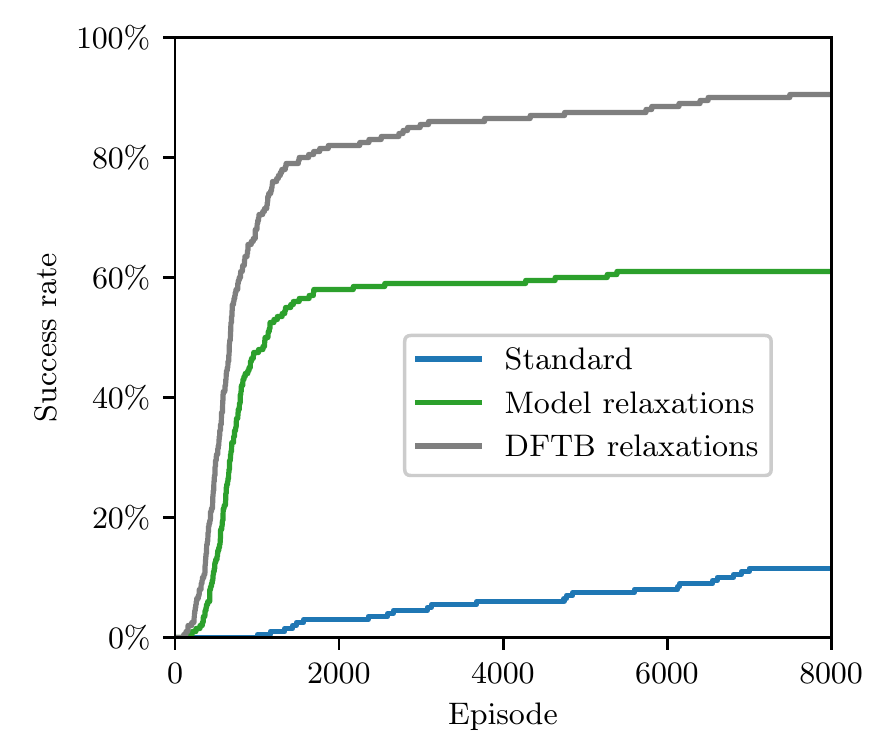}
  \caption{Success curves based on 200 restarts. The standard ASLA (blue
    curve) is outperformed by ASLA augmented with model relaxations
    (green curve), without any additional energy evaluations. The grey
    curve shows the performance when DFTB relaxations are used
    in place of the model relaxations, imitating a perfectly trained
    model. }
  \label{benzene_results}    
\end{figure}

\begin{figure}
  \includegraphics[width=0.45\textwidth]{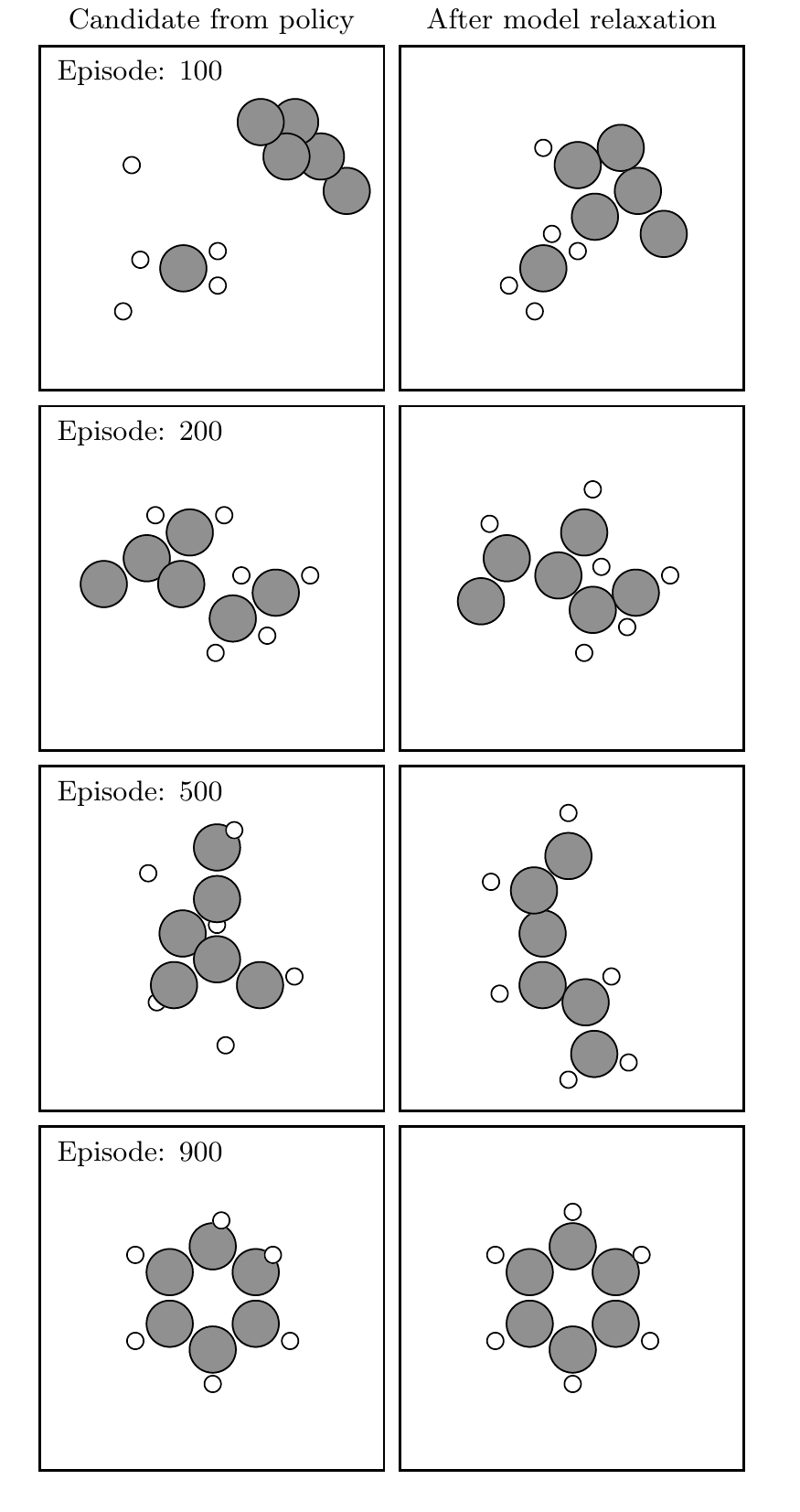}      
  \caption{The structures built according to the $Q$-value directed policy (left)
    and their appearance being relaxed in the model and snapped to the grid  (right)
    for different stages of a search. The model potential quickly
    learns to locally optimize the structure that ASLA produces. In
    turn, ASLA learns more quickly to build stable structures
    (cf.\ Fig.\ \protect\ref{benzene_results}).
  }
  \label{benzene_episodes}    
\end{figure}

The performance of ASLA is gauged by restarting the structural search
a great number of times with no data in the memory and with random
initialization of the CNN. By compiling as a
function of episode count, the share of search runs that have found
the benzene molecule, we arrive at the success curves displayed in
Fig.\ \ref{benzene_results}. The blue curve shows the performance
of standard ASLA without model relaxations, while the green curve
shows the performance when the model relaxations are included. The
here-proposed ASLA with model relaxations outperforms the standard
ASLA by a wide margin. While standard ASLA needs $\approx7000$
episodes for 10\% of the restarts to identify benzene, ASLA with model
relaxations attains this after less than 1000 episodes.

The green success curve for the model-relaxed ASLA shows, however, one
convergence issue. The fact that it levels off at about 60\% success
after about 2000 episodes means that the 40\% of the restarts that have not found
benzene within the first 2000 episodes stand a small chance of doing
so in the remaining 6000 episodes of the runs. To probe
the origin of this, we conducted a set of search runs, in which the
relaxations of the policy-built candidate structures were done in the
full DFTB energy landscape rather than with the on-the-fly learnt
surrogate energy landscape. The grey curve in
Fig.\ \ref{benzene_results} shows the resulting success curve, which
evidences that when relaxing candidates in the DFTB energy landscape,
the stagnation almost vanishes. About 80\% of the restarts have found
benzene after 2000 episodes, and in the following 6000 episodes about
half of the remaining restarts do so.

We consequently attribute the stagnation of the model-relaxed ASLA to
errors in the surrogate energy model. Note, however, that the DFTB
relaxation scheme uses several order of magnitudes more DFTB energy
evaluations per episode, than does the model relaxation scheme, where
one episode corresponds to one DFTB single-point energy evaluation. This
renders the model relaxation scheme far superior when more refined and
computationally expensive DFT or quantum
chemistry methods are used as the total energy expression.

Possible solutions to the stagnation issue could be the introduction
of more advanced model energy expressions than the presently used one,
or the development of schemes to reset the surrogate model upon
detection of stagnation.  Note that a single restart will not know if
it has ceased to find better structures because it has identified the
global minimum energy structure or because it uses an insufficiently
accurate model for the relaxation and finds a higher lying energy
structure. In practice, when searching for an unknown global minimum
energy structure, all restarts would therefore need to have their
model reset once they consistently produce the same best structure
over and over again.

Having seen that ASLA improves significantly upon adding the presently
proposed model-relaxation, it is instructive to inspect the degree of
relaxation as a function of reinforcement episodes.  Figure
\ref{benzene_episodes} presents one such study for a randomly chosen
restart. It shows the policy-built structure and the corresponding
model-relaxed structure after 100, 200, 500, and 900 episodes. After
100 and 200 episodes, the agents CNN has not
developed sufficiently yet as to consistently build molecules with
chemical meaningful coordinations. However, the surrogate energy
model has learnt enough to provide for relaxation of the atoms into
having more proper interatomic distances. After 500 episodes, the policy-built
structure still lacks a bit on C-C coordination and C-H bond lengths
and it leaves some H as isolated atoms. All of these deficiencies are
remedied by the model-relaxation and by 900 episodes, the agents
starts to know how to build benzene in need of only very minor bond
adjustments, mainly pertaining to the large grid spacing used. Note,
that the CNN benefits from the improved training examples provided by
the model relaxations, and learns to build reasonable structures
faster than in the standard ASLA scheme.

\begin{figure}
  \includegraphics[width=0.45\textwidth]{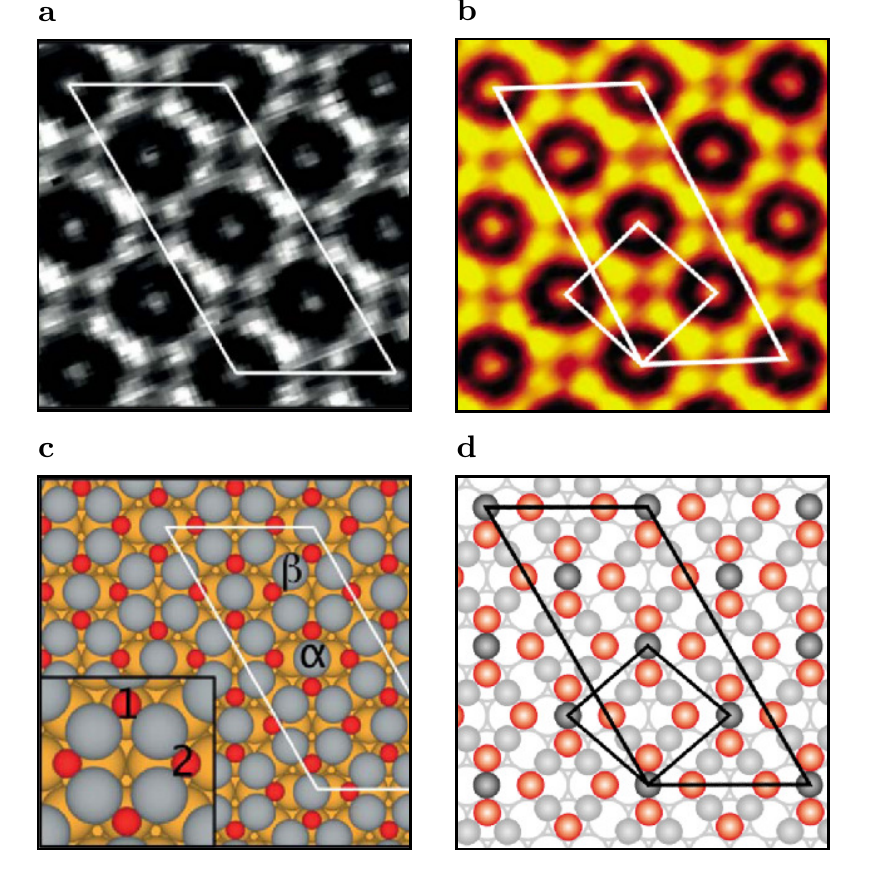}
  \caption{The experimental STM images and conjectured atomistic
structures. \textbf{a} and \textbf{c} reprinted with permission from Martin \emph{et al} The Journal of Physical Chemistry C \textbf{118}, 15324 (2014). Copyright (2020) American Chemical Society. 
\textbf{b} and \textbf{d} adapted from Ref.\ \protect\onlinecite{schnadt_prb_2009}.
}
    \label{fig:stm}
\end{figure}

\begin{figure}
  \includegraphics[scale=0.6]{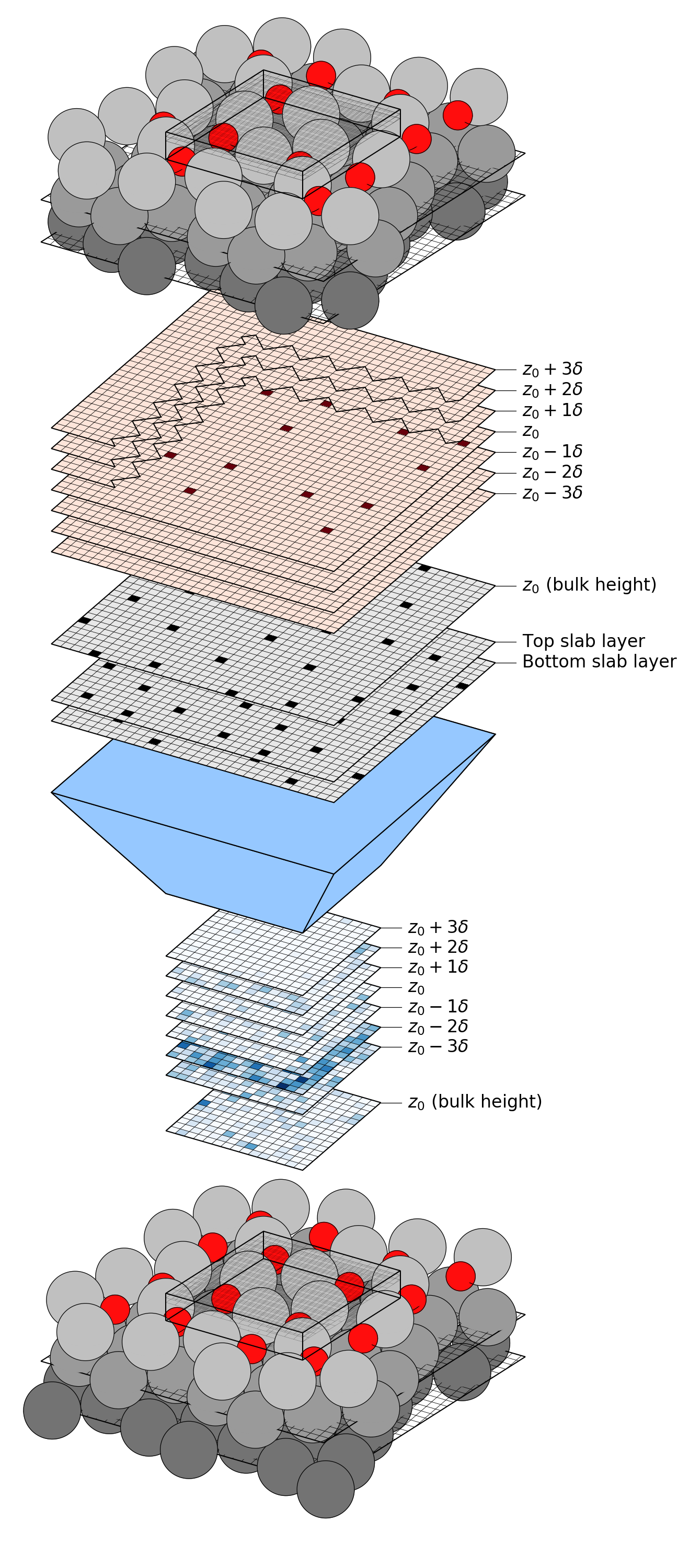}
  \caption{The ASLA setup.
Starting from above, an incomplete surface oxide structure on a
two-layer Ag(111) template
is shown. Red spheres are oxygen atoms, and grey 
spheres are silver atoms. The transparent box highlights the
$c(4\times 8)$ unit cell. 
The two lowest layers of Ag atoms, two Ag(111) layers,
are part of the template, while the upper most Ag atoms and all O atoms
are being placed by ASLA. The convolutional neural network is fed with
seven layers of allowed O positions (the brown grids) and with three
layers of possible Ag positions (the grey grids). It outputs $Q$-values
for all available
oxygen atomic positions and for silver positions in the upper layer. In
the final lower structure, an extra O atom has been added at the
position of highest $Q$-value following a greedy policy.}
  \label{Ag111_ASLA_setup}  
\end{figure}

\section{Ag(111) oxide surface}

\begin{figure}[tb]
  \centering
  \includegraphics[width=0.45\textwidth]{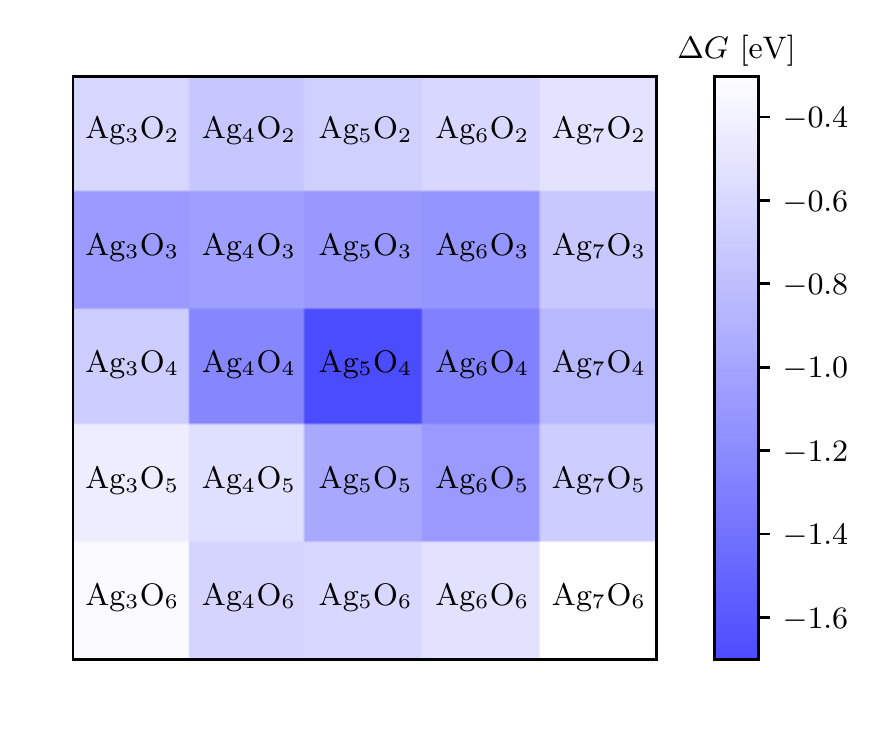}
  \caption{Free energy, $\Delta G$ evaluated at $\Delta\mu_\mathrm{O}=-0.5$ eV, for
    the most stable Ag(111)-$c(4\times 8)$-Ag$_X$O$_Y$ surface oxides found by ASLA.}
  \label{fig:raster}
\end{figure}

We now turn to apply ASLA with model relaxation to an outstanding problem in
materials science.  Specifically, we will be concerned with the
oxidation of the Ag(111)-surface, which has been shown to exhibit a
rich variety of stoichiometries and phases during growth of surface
oxide layers. Schnadt and co-workers \cite{schnadt_prb_2009,martin_j_phys_chem_c_2014} have reported scanning
tunneling microscopy (STM) topographs of a number of such phases, one
of which, the $c(4\times 8)$-phase, is reproduced in
Fig.\ \ref{fig:stm}.  To the best of our knowledge, no systematic
structural search has been carried out for this particular 
surface oxide phase on Ag(111), yet in the
original paper a structural model was put forward \cite{schnadt_prb_2009}. By having
ASLA perform the structural search given the experimentally determined
surface unit cell, but covering a large range of Ag-O stoichiometries
for the surface oxide, we confirm below the originally conjectured
model in a set of ASLA runs.

In our search for the Ag(111)-$c(4\times 8)$ silver oxide phase, ASLA is
used to identify the best possible structure of a mono-layer silver
oxide on top of a bulk truncated Ag(111) template. For the surface
oxide, it
is assumed that all Ag atoms occupy positions whose heights coincide
with that of an extra unrelaxed Ag(111) layer, $z_0$, while all O atoms occupy some of seven
different heights evenly distributed around such a Ag(111) layer, i.e.\
ranging from 1.6 to 3.1 {\AA} above the Ag(111) template. For
the in-plane positions of the Ag and O atoms, any values on a 2D
grid with grid spacing 0.255 {\AA} are allowed. Figure
\ref{Ag111_ASLA_setup} depicts these discretized positions as they are
fed into the neural network. The seven brown layers hold the O
atoms, while the three grey layers hold the Ag atoms. As a template
of preplaced Ag atoms, the bottom two grey layers are prepopulated with
two Ag(111) layers. The CNN, shown schematically
as the blue polygon in Fig.\ \ref{Ag111_ASLA_setup}, outputs $Q$-values
for all oxygen layers and for the upper Ag layer as shown by the bluish
raster plots in the figure. The network does not need to output
any $Q$-values for the lower two Ag layers, since these are already fully
occupied.

During the model relaxation the silver atoms are constrained to move in the
$xy$-plane, while the oxygen atoms are constrained to move freely
between $z=1.6$ {\AA} and $z = 3.1$ {\AA}, which does not represent any
further approximation given that the atomic position are snapped to the
grid heights within these bounds at the end of the building phase.

The evaluation phase is conducted using DFT.
The structures are handled with the atomic simulation
environment (ASE) \cite{ase} and the DFT based total energies are
evaluated with the grid-based projector augmented wave package
{\sc gpaw} \cite{gpaw2}. For the exchange-correlation functional, the
Perdew-Burke-Ernzerhof (PBE) expression is used \cite{pbe}.

ASLA was used to build Ag$_X$O$_Y$ structures on top of Ag(111)-$c(4\times 8)$
for $3\le X\le 7$ and $2\le Y\le 6$. For each combination of $(X,Y)$ ASLA
was restarted at least 10 times until the best structure found agreed
in at least 5 of the restarts, except for some of the
Ag$_7$O$_Y$ runs, where only a few restarts agreed on the
best structure before the allocated resources were spent. Once identified by
ASLA, the best surface oxide structure ever found for each restart was transferred to
5 layered Ag(111) slabs and relaxed unconstrained. The structure and stability of
these surface oxide models are discussed in the following.

\begin{figure}[tb]
  \centering
  \includegraphics[width=0.45\textwidth]{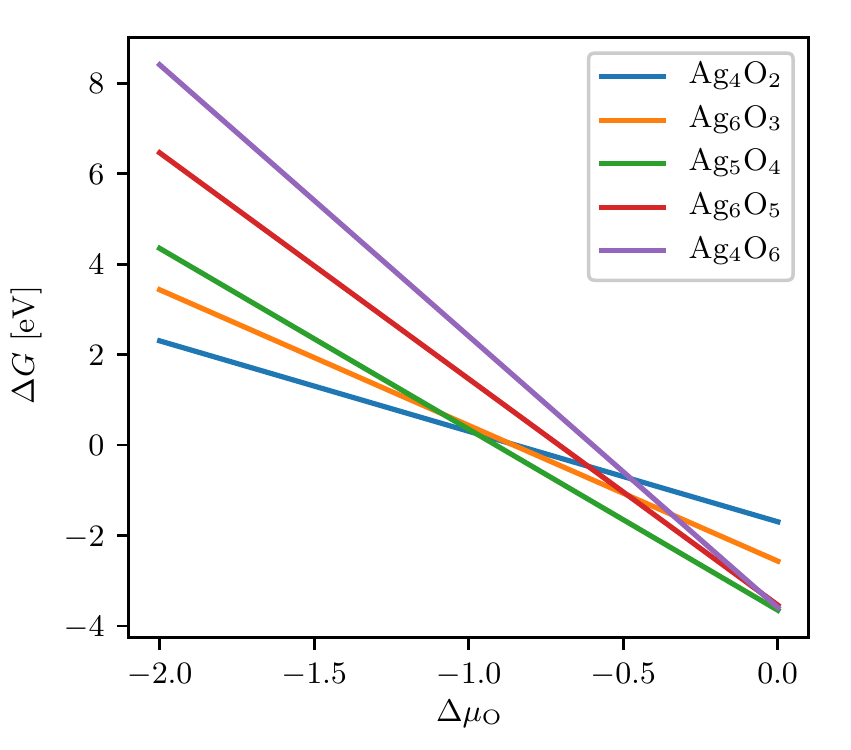}
  \caption{Free energy, $\Delta G$ as a function of $\Delta\mu_\mathrm{O}$ for
the most stable Ag(111)-$c(4\times 8)$-Ag$_X$O$_Y$ surface oxides found by ASLA for $Y=2, 3, 4, 5$ and $6$
when varying $X$ from 3 to 7.}
  \label{fig:phase}
\end{figure}

\begin{figure*}
  \centering
  \includegraphics{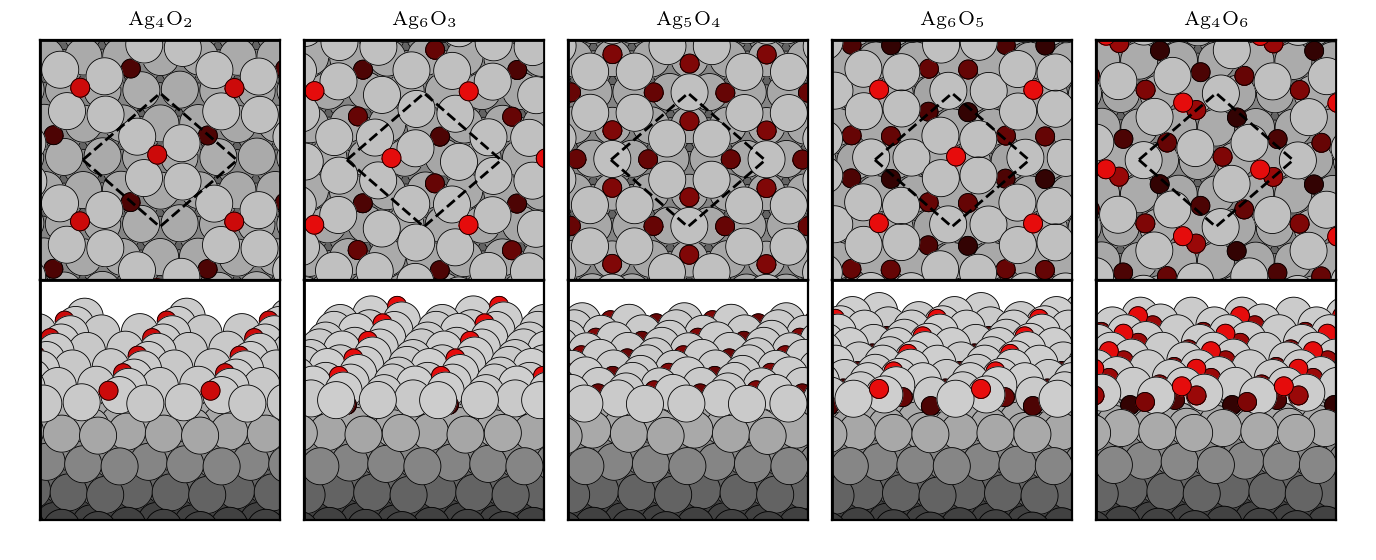}
  \caption{Top and side view of the most stable Ag$_X$O$_Y$ structures
    for each value of $Y$ investigated in this work.}
  \label{fig:structures}
\end{figure*}

We compare the different stoichiometries by their Gibbs free energy
according to \cite{reuter_prb_2003}:
\begin{equation}
  \label{eq: free energy}
\Delta G (T,p)=  E^\mathrm{DFT}-E^\mathrm{DFT}_\mathrm{slab} - X\mu_{\text{Ag}} -Y\mu_{\text{O}}(T,p),
\end{equation}
where $E^\mathrm{DFT}$ is the DFT energy of the full structure, $E^\mathrm{DFT}_\mathrm{slab}$ is the
DFT energy of the Ag(111)-$c(4\times 8)$ five layer slab without the surface oxide, and where 
$\mu_{\text{Ag}}$ and $\mu_{\text{O}}$ are the chemical potentials of silver and oxygen, respectively.
Note, that we neglect the small vibrational and configurational contributions to the
Gibbs free energy of the surface structures. The silver chemical potential, $\mu_{\text{Ag}}$, is taken to be the chemical
potential of a silver atom situated in the bulk position, calculated as the difference per Ag atom
of 6 and 5 layer thick Ag(111) slabs. The
oxygen chemical potential is a function of the di-oxygen pressure, $p$, and the temperature, $T$ \cite{reuter_prb_2003}:
\begin{equation}
  \label{eq: mu O}
  \mu_\mathrm{O}(T,p) = \frac{1}{2} \left( E_{\mathrm{O}_2}^\mathrm{DFT} + \tilde{\mu}_\mathrm{O_2}(T,p^0) + k_B T \ln \left( \frac{p}{p^0}\right) \right)
\end{equation}
where $E_{\mathrm{O}_2}^\mathrm{DFT}$ is the DFT total energy of the di-oxygen
molecule, $\tilde{\mu}_\mathrm{O_2}(T,p^0)$ is the translational,
rotational, and vibrational contributions to the free energy of an O$_2$ gas at a reference
pressure, $p_0$, and $k_B$ is Boltzmanns constant. It is seen that the
absolute value of $\mu_\mathrm{O}(T,p)$ depends on the specific
computational settings through $E_{\mathrm{O}_2}^\mathrm{DFT}$. To circumvent
that, it is convenient to quote only the $(T,p)$-dependent part of the chemical
potential:
\begin{equation}
\Delta\mu_{\text{O}}(T,p)=\mu_O(T,p) - \frac{1}{2}E_{\mathrm{O}_2}^\mathrm{DFT}
\end{equation}
whenever a chemical potential is specified.

With the thermodynamic considerations in place, it is now possible to
compare the stability of the most stable structures found by ASLA. Figure
\ref{fig:raster} does so in the form of a raster plot of the free energy, $\Delta G$,
evaluated at an oxygen chemical potential of $\Delta\mu_\mathrm{O}=-0.5$
 eV, corresponding approximately to ambient conditions, i.e.\ a pressure
of 1 atm and a temperature of 300 K \cite{reuter_prb_2003}.
The plot shows a clear optimal stability for a surface
oxide of Ag$_5$O$_4$ stoichiometry, and it brings evidence that
sufficient variation in the stoichiometry has been considered to call
this the thermodynamic most stable state at this chemical
potential for oxygen.

In Fig.\ \ref{fig:phase} a diagram of the free energy as a function of
$\Delta\mu_{\text{O}}$ is shown. The diagram builds on the most stable
Ag$_X$O$_Y$ structure for every considered value of $Y$, i.e.\
Ag$_4$O$_2$, Ag$_6$O$_3$, Ag$_5$O$_4$, Ag$_6$O$_5$, and
Ag$_4$O$_6$ surface oxides. The diagram shows that over a wide
range of chemical potential for oxygen, the Ag$_5$O$_4$ surface oxide
remains the most stable.

\begin{figure}[tb]
  \centering
 \includegraphics{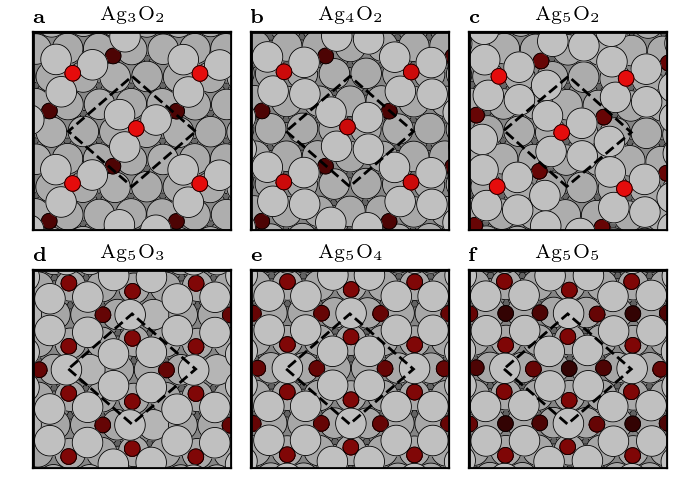}
 \caption{Structures with similar motifs found for different
   stoichometries. \textbf{a}-\textbf{c} show network-like structures
   for Ag$_3$O$_2$, Ag$_4$O$_2$ and Ag$_5$O$_2$. \textbf{d}-\textbf{f}
   show structures for Ag$_5$O$_3$, Ag$_5$O$_4$ and Ag$_5$O$_5$. Here,
   identical motifs are found for the Ag-atoms, while the oxygen sites
   are filled as more oxygen are introduced. Note that all structures
   above are found in separate, independent ASLA runs.  }
  \label{fig:other_structures}
\end{figure}

The structures leading to the free energy diagram in
Fig.\ \ref{fig:phase} are shown in Fig.\ \ref{fig:structures}. They
expose a rich variety of chemical bonding motifs involving the Ag and
O atoms within the surface oxide.
Inspecting the structures in Fig.\ \ref{fig:structures}, the first one
appears network type with voids, the next one resembles the first, but
with reduced Ag$_3$ islands in the voids, while the remaining three
structures appear to have both Ag and and O atoms highly dispersed in
the surface oxide layer. As the oxygen content increases from left to
right in the figure, ASLA eventually identifies the need for
including O$_2$ motifs when tasked with accommodating a large amount of
oxygen within the surface oxide.

The large diversity in the \emph{optimal} structures shown in
Fig.\ \ref{fig:structures} testifies to ASLAs ability to identify a
highly diverse set of chemically meaningful structures as the
stoichiometry is varied. Conversely, collecting series of some of the
\emph{sub-optimal} structures found, it can be realized that ASLA
often finds the same structural skeleton for different
stoichiometries. This is illustrated in
Fig.\ \ref{fig:other_structures}a-c, where it is seen that ASLA in
independent searches for Ag$_3$O$_2$ and Ag$_5$O$_2$ stoichiometries
finds again the optimal Ag$_4$O$_2$ structure 
only with one Ag atom removed or
added. Likewise, Fig.\ \ref{fig:other_structures}d-f shows how
Ag$_5$O$_3$ and Ag$_5$O$_5$ structures are identified by ASLA that
either lack one O or contain an extra O compared to the optimal
Ag$_5$O$_4$. A full account of all structures
found is given in the supplementary material.

\begin{figure}
  \includegraphics{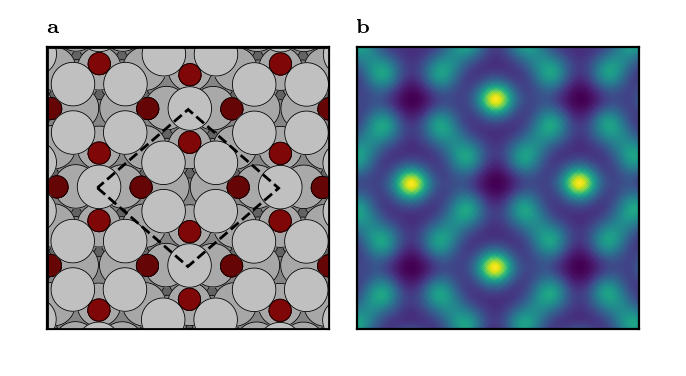}
  \caption{\textbf{a} The Ag$_5$O$_4$ structure and \textbf{b} the
    corresponding simulated STM image. The simulated STM agrees well
    with the experimental STM image (see Fig. \ref{fig:stm}
    \textbf{a} and \textbf{b}). }
    \label{fig:sim_stm}
\end{figure}

Focusing on the Ag$_5$O$_4$ surface oxide that we identify as the
preferred structure, we note that ASLA has indeed found the same
structure that was proposed by Schnadt \emph{et al} \cite{schnadt_prb_2009}. In
Fig.\ \ref{fig:sim_stm} we present a Tersoff-Hermann type simulated
STM topograph, that indeed matches the experimental STM one shown
in Fig.\ \ref{fig:stm}. We thus conclude that ASLA with
very little input (i.e. the surface unit cell and some choice for discretization
of space) is capable of deducing the structure of a complicated surface
oxide. It does so without requiring any human input of expected structural
motifs or other presumed knowledge of chemical bonding within the 
surface oxide.

\section{Conclusion}
We have augmented ASLA with relaxations in a model potential,
dramatically increasing the performance. The improvements come with no
need for extra electronic-structure energy evaluations, and little overhead due to training
the model and the relaxation procedure itself. We applied the improved
ASLA to a Ag(111) oxide surface, where a proposed structure was
confirmed by a thorough search.

\section{Acknowledgements}
We acknowledge support from VILLUM FONDEN (Investigator grant, Project No.\ 16562).

\bibliographystyle{apsrev4-1}
\bibliography{references}

\begin{thebibliography}{61}%
\makeatletter
\providecommand \@ifxundefined [1]{%
 \@ifx{#1\undefined}
}%
\providecommand \@ifnum [1]{%
 \ifnum #1\expandafter \@firstoftwo
 \else \expandafter \@secondoftwo
 \fi
}%
\providecommand \@ifx [1]{%
 \ifx #1\expandafter \@firstoftwo
 \else \expandafter \@secondoftwo
 \fi
}%
\providecommand \natexlab [1]{#1}%
\providecommand \enquote  [1]{``#1''}%
\providecommand \bibnamefont  [1]{#1}%
\providecommand \bibfnamefont [1]{#1}%
\providecommand \citenamefont [1]{#1}%
\providecommand \href@noop [0]{\@secondoftwo}%
\providecommand \href [0]{\begingroup \@sanitize@url \@href}%
\providecommand \@href[1]{\@@startlink{#1}\@@href}%
\providecommand \@@href[1]{\endgroup#1\@@endlink}%
\providecommand \@sanitize@url [0]{\catcode `\\12\catcode `\$12\catcode
  `\&12\catcode `\#12\catcode `\^12\catcode `\_12\catcode `\%12\relax}%
\providecommand \@@startlink[1]{}%
\providecommand \@@endlink[0]{}%
\providecommand \url  [0]{\begingroup\@sanitize@url \@url }%
\providecommand \@url [1]{\endgroup\@href {#1}{\urlprefix }}%
\providecommand \urlprefix  [0]{URL }%
\providecommand \Eprint [0]{\href }%
\providecommand \doibase [0]{http://dx.doi.org/}%
\providecommand \selectlanguage [0]{\@gobble}%
\providecommand \bibinfo  [0]{\@secondoftwo}%
\providecommand \bibfield  [0]{\@secondoftwo}%
\providecommand \translation [1]{[#1]}%
\providecommand \BibitemOpen [0]{}%
\providecommand \bibitemStop [0]{}%
\providecommand \bibitemNoStop [0]{.\EOS\space}%
\providecommand \EOS [0]{\spacefactor3000\relax}%
\providecommand \BibitemShut  [1]{\csname bibitem#1\endcsname}%
\let\auto@bib@innerbib\@empty
\bibitem [{\citenamefont {Oganov}\ \emph {et~al.}(2019)\citenamefont {Oganov},
  \citenamefont {Pickard}, \citenamefont {Zhu},\ and\ \citenamefont
  {Needs}}]{materialrs}%
  \BibitemOpen
  \bibfield  {author} {\bibinfo {author} {\bibfnamefont {A.~R.}\ \bibnamefont
  {Oganov}}, \bibinfo {author} {\bibfnamefont {C.~J.}\ \bibnamefont {Pickard}},
  \bibinfo {author} {\bibfnamefont {Q.}~\bibnamefont {Zhu}}, \ and\ \bibinfo
  {author} {\bibfnamefont {R.~J.}\ \bibnamefont {Needs}},\ }\href {\doibase
  10.1038/s41578-019-0101-8} {\bibfield  {journal} {\bibinfo  {journal} {Nature
  Reviews Materials}\ }\textbf {\bibinfo {volume} {4}},\ \bibinfo {pages} {331}
  (\bibinfo {year} {2019})}\BibitemShut {NoStop}%
\bibitem [{\citenamefont {Kirkpatrick}\ \emph {et~al.}(1983)\citenamefont
  {Kirkpatrick}, \citenamefont {Gelatt},\ and\ \citenamefont
  {Vecchi}}]{sim_an}%
  \BibitemOpen
  \bibfield  {author} {\bibinfo {author} {\bibfnamefont {S.}~\bibnamefont
  {Kirkpatrick}}, \bibinfo {author} {\bibfnamefont {C.~D.}\ \bibnamefont
  {Gelatt}}, \ and\ \bibinfo {author} {\bibfnamefont {M.~P.}\ \bibnamefont
  {Vecchi}},\ }\href {\doibase 10.1126/science.220.4598.671} {\bibfield
  {journal} {\bibinfo  {journal} {Science}\ }\textbf {\bibinfo {volume}
  {220}},\ \bibinfo {pages} {671} (\bibinfo {year} {1983})}\BibitemShut
  {NoStop}%
\bibitem [{\citenamefont {Wales}\ and\ \citenamefont {Doye}(1997)}]{basin_hop}%
  \BibitemOpen
  \bibfield  {author} {\bibinfo {author} {\bibfnamefont {D.~J.}\ \bibnamefont
  {Wales}}\ and\ \bibinfo {author} {\bibfnamefont {J.~P.~K.}\ \bibnamefont
  {Doye}},\ }\href {\doibase 10.1021/jp970984n} {\bibfield  {journal} {\bibinfo
   {journal} {The Journal of Physical Chemistry A}\ }\textbf {\bibinfo {volume}
  {101}},\ \bibinfo {pages} {5111} (\bibinfo {year} {1997})}\BibitemShut
  {NoStop}%
\bibitem [{\citenamefont {Hartke}(1993)}]{GA1}%
  \BibitemOpen
  \bibfield  {author} {\bibinfo {author} {\bibfnamefont {B.}~\bibnamefont
  {Hartke}},\ }\href {\doibase 10.1021/j100141a013} {\bibfield  {journal}
  {\bibinfo  {journal} {The Journal of Physical Chemistry}\ }\textbf {\bibinfo
  {volume} {97}},\ \bibinfo {pages} {9973} (\bibinfo {year}
  {1993})}\BibitemShut {NoStop}%
\bibitem [{\citenamefont {Deaven}\ and\ \citenamefont {Ho}(1995)}]{GA2}%
  \BibitemOpen
  \bibfield  {author} {\bibinfo {author} {\bibfnamefont {D.~M.}\ \bibnamefont
  {Deaven}}\ and\ \bibinfo {author} {\bibfnamefont {K.~M.}\ \bibnamefont
  {Ho}},\ }\href {\doibase 10.1103/PhysRevLett.75.288} {\bibfield  {journal}
  {\bibinfo  {journal} {Phys. Rev. Lett.}\ }\textbf {\bibinfo {volume} {75}},\
  \bibinfo {pages} {288} (\bibinfo {year} {1995})}\BibitemShut {NoStop}%
\bibitem [{\citenamefont {Oganov}\ and\ \citenamefont {Glass}(2006)}]{GA3}%
  \BibitemOpen
  \bibfield  {author} {\bibinfo {author} {\bibfnamefont {A.~R.}\ \bibnamefont
  {Oganov}}\ and\ \bibinfo {author} {\bibfnamefont {C.~W.}\ \bibnamefont
  {Glass}},\ }\href {\doibase 10.1063/1.2210932} {\bibfield  {journal}
  {\bibinfo  {journal} {The Journal of Chemical Physics}\ }\textbf {\bibinfo
  {volume} {124}},\ \bibinfo {pages} {244704} (\bibinfo {year}
  {2006})}\BibitemShut {NoStop}%
\bibitem [{\citenamefont {Vilhelmsen}\ and\ \citenamefont
  {Hammer}(2012)}]{GA4}%
  \BibitemOpen
  \bibfield  {author} {\bibinfo {author} {\bibfnamefont {L.~B.}\ \bibnamefont
  {Vilhelmsen}}\ and\ \bibinfo {author} {\bibfnamefont {B.}~\bibnamefont
  {Hammer}},\ }\href {\doibase 10.1103/PhysRevLett.108.126101} {\bibfield
  {journal} {\bibinfo  {journal} {Phys. Rev. Lett.}\ }\textbf {\bibinfo
  {volume} {108}},\ \bibinfo {pages} {126101} (\bibinfo {year}
  {2012})}\BibitemShut {NoStop}%
\bibitem [{\citenamefont {Vilhelmsen}\ and\ \citenamefont
  {Hammer}(2014)}]{GA5}%
  \BibitemOpen
  \bibfield  {author} {\bibinfo {author} {\bibfnamefont {L.~B.}\ \bibnamefont
  {Vilhelmsen}}\ and\ \bibinfo {author} {\bibfnamefont {B.}~\bibnamefont
  {Hammer}},\ }\href {\doibase 10.1063/1.4886337} {\bibfield  {journal}
  {\bibinfo  {journal} {The Journal of Chemical Physics}\ }\textbf {\bibinfo
  {volume} {141}},\ \bibinfo {pages} {044711} (\bibinfo {year}
  {2014})}\BibitemShut {NoStop}%
\bibitem [{\citenamefont {Ishikawa}\ \emph {et~al.}(2019)\citenamefont
  {Ishikawa}, \citenamefont {Miyake},\ and\ \citenamefont {Shimizu}}]{GA6}%
  \BibitemOpen
  \bibfield  {author} {\bibinfo {author} {\bibfnamefont {T.}~\bibnamefont
  {Ishikawa}}, \bibinfo {author} {\bibfnamefont {T.}~\bibnamefont {Miyake}}, \
  and\ \bibinfo {author} {\bibfnamefont {K.}~\bibnamefont {Shimizu}},\ }\href
  {\doibase 10.1103/PhysRevB.100.174506} {\bibfield  {journal} {\bibinfo
  {journal} {Phys. Rev. B}\ }\textbf {\bibinfo {volume} {100}},\ \bibinfo
  {pages} {174506} (\bibinfo {year} {2019})}\BibitemShut {NoStop}%
\bibitem [{\citenamefont {Schmidt}\ \emph {et~al.}(2019)\citenamefont
  {Schmidt}, \citenamefont {Marques}, \citenamefont {Botti},\ and\
  \citenamefont {Marques}}]{Schmidt_review_2019}%
  \BibitemOpen
  \bibfield  {author} {\bibinfo {author} {\bibfnamefont {J.}~\bibnamefont
  {Schmidt}}, \bibinfo {author} {\bibfnamefont {M.~R.~G.}\ \bibnamefont
  {Marques}}, \bibinfo {author} {\bibfnamefont {S.}~\bibnamefont {Botti}}, \
  and\ \bibinfo {author} {\bibfnamefont {M.~A.~L.}\ \bibnamefont {Marques}},\
  }\href {\doibase 10.1038/s41524-019-0221-0} {\bibfield  {journal} {\bibinfo
  {journal} {npj Computational Materials}\ }\textbf {\bibinfo {volume} {5}},\
  \bibinfo {pages} {83} (\bibinfo {year} {2019})}\BibitemShut {NoStop}%
\bibitem [{\citenamefont {Bart\'ok}\ \emph {et~al.}(2010)\citenamefont
  {Bart\'ok}, \citenamefont {Payne}, \citenamefont {Kondor},\ and\
  \citenamefont {Cs\'anyi}}]{K1}%
  \BibitemOpen
  \bibfield  {author} {\bibinfo {author} {\bibfnamefont {A.~P.}\ \bibnamefont
  {Bart\'ok}}, \bibinfo {author} {\bibfnamefont {M.~C.}\ \bibnamefont {Payne}},
  \bibinfo {author} {\bibfnamefont {R.}~\bibnamefont {Kondor}}, \ and\ \bibinfo
  {author} {\bibfnamefont {G.}~\bibnamefont {Cs\'anyi}},\ }\href {\doibase
  10.1103/PhysRevLett.104.136403} {\bibfield  {journal} {\bibinfo  {journal}
  {Phys. Rev. Lett.}\ }\textbf {\bibinfo {volume} {104}},\ \bibinfo {pages}
  {136403} (\bibinfo {year} {2010})}\BibitemShut {NoStop}%
\bibitem [{\citenamefont {Rupp}\ \emph {et~al.}(2012)\citenamefont {Rupp},
  \citenamefont {Tkatchenko}, \citenamefont {M\"uller},\ and\ \citenamefont
  {von Lilienfeld}}]{K2}%
  \BibitemOpen
  \bibfield  {author} {\bibinfo {author} {\bibfnamefont {M.}~\bibnamefont
  {Rupp}}, \bibinfo {author} {\bibfnamefont {A.}~\bibnamefont {Tkatchenko}},
  \bibinfo {author} {\bibfnamefont {K.-R.}\ \bibnamefont {M\"uller}}, \ and\
  \bibinfo {author} {\bibfnamefont {O.~A.}\ \bibnamefont {von Lilienfeld}},\
  }\href {\doibase 10.1103/PhysRevLett.108.058301} {\bibfield  {journal}
  {\bibinfo  {journal} {Phys. Rev. Lett.}\ }\textbf {\bibinfo {volume} {108}},\
  \bibinfo {pages} {058301} (\bibinfo {year} {2012})}\BibitemShut {NoStop}%
\bibitem [{\citenamefont {Bart\'ok}\ \emph {et~al.}(2013)\citenamefont
  {Bart\'ok}, \citenamefont {Kondor},\ and\ \citenamefont {Cs\'anyi}}]{K3}%
  \BibitemOpen
  \bibfield  {author} {\bibinfo {author} {\bibfnamefont {A.~P.}\ \bibnamefont
  {Bart\'ok}}, \bibinfo {author} {\bibfnamefont {R.}~\bibnamefont {Kondor}}, \
  and\ \bibinfo {author} {\bibfnamefont {G.}~\bibnamefont {Cs\'anyi}},\ }\href
  {\doibase 10.1103/PhysRevB.87.184115} {\bibfield  {journal} {\bibinfo
  {journal} {Phys. Rev. B}\ }\textbf {\bibinfo {volume} {87}},\ \bibinfo
  {pages} {184115} (\bibinfo {year} {2013})}\BibitemShut {NoStop}%
\bibitem [{\citenamefont {Behler}\ and\ \citenamefont
  {Parrinello}(2007)}]{NN1}%
  \BibitemOpen
  \bibfield  {author} {\bibinfo {author} {\bibfnamefont {J.}~\bibnamefont
  {Behler}}\ and\ \bibinfo {author} {\bibfnamefont {M.}~\bibnamefont
  {Parrinello}},\ }\href {\doibase 10.1103/PhysRevLett.98.146401} {\bibfield
  {journal} {\bibinfo  {journal} {Phys. Rev. Lett.}\ }\textbf {\bibinfo
  {volume} {98}},\ \bibinfo {pages} {146401} (\bibinfo {year}
  {2007})}\BibitemShut {NoStop}%
\bibitem [{\citenamefont {Sch\"{u}tt}\ \emph {et~al.}(2018)\citenamefont
  {Sch\"{u}tt}, \citenamefont {Sauceda}, \citenamefont {Kindermans},
  \citenamefont {Tkatchenko},\ and\ \citenamefont {M\"{u}ller}}]{NN2}%
  \BibitemOpen
  \bibfield  {author} {\bibinfo {author} {\bibfnamefont {K.~T.}\ \bibnamefont
  {Sch\"{u}tt}}, \bibinfo {author} {\bibfnamefont {H.~E.}\ \bibnamefont
  {Sauceda}}, \bibinfo {author} {\bibfnamefont {P.-J.}\ \bibnamefont
  {Kindermans}}, \bibinfo {author} {\bibfnamefont {A.}~\bibnamefont
  {Tkatchenko}}, \ and\ \bibinfo {author} {\bibfnamefont {K.-R.}\ \bibnamefont
  {M\"{u}ller}},\ }\href {\doibase 10.1063/1.5019779} {\bibfield  {journal}
  {\bibinfo  {journal} {The Journal of Chemical Physics}\ }\textbf {\bibinfo
  {volume} {148}},\ \bibinfo {pages} {241722} (\bibinfo {year}
  {2018})}\BibitemShut {NoStop}%
\bibitem [{\citenamefont {Smith}\ \emph {et~al.}(2019)\citenamefont {Smith},
  \citenamefont {Nebgen}, \citenamefont {Zubatyuk}, \citenamefont {Lubbers},
  \citenamefont {Devereux}, \citenamefont {Barros}, \citenamefont {Tretiak},
  \citenamefont {Isayev},\ and\ \citenamefont {Roitberg}}]{NN3}%
  \BibitemOpen
  \bibfield  {author} {\bibinfo {author} {\bibfnamefont {J.~S.}\ \bibnamefont
  {Smith}}, \bibinfo {author} {\bibfnamefont {B.~T.}\ \bibnamefont {Nebgen}},
  \bibinfo {author} {\bibfnamefont {R.}~\bibnamefont {Zubatyuk}}, \bibinfo
  {author} {\bibfnamefont {N.}~\bibnamefont {Lubbers}}, \bibinfo {author}
  {\bibfnamefont {C.}~\bibnamefont {Devereux}}, \bibinfo {author}
  {\bibfnamefont {K.}~\bibnamefont {Barros}}, \bibinfo {author} {\bibfnamefont
  {S.}~\bibnamefont {Tretiak}}, \bibinfo {author} {\bibfnamefont
  {O.}~\bibnamefont {Isayev}}, \ and\ \bibinfo {author} {\bibfnamefont {A.~E.}\
  \bibnamefont {Roitberg}},\ }\href {\doibase 10.1038/s41467-019-10827-4}
  {\bibfield  {journal} {\bibinfo  {journal} {Nature Communications}\ }\textbf
  {\bibinfo {volume} {10}},\ \bibinfo {pages} {2903} (\bibinfo {year}
  {2019})}\BibitemShut {NoStop}%
\bibitem [{\citenamefont {Sch\"{u}tt}\ \emph {et~al.}(2019)\citenamefont
  {Sch\"{u}tt}, \citenamefont {Gastegger}, \citenamefont {Tkatchenko},
  \citenamefont {M\"{u}ller},\ and\ \citenamefont {Maurer}}]{NN4}%
  \BibitemOpen
  \bibfield  {author} {\bibinfo {author} {\bibfnamefont {K.~T.}\ \bibnamefont
  {Sch\"{u}tt}}, \bibinfo {author} {\bibfnamefont {M.}~\bibnamefont
  {Gastegger}}, \bibinfo {author} {\bibfnamefont {A.}~\bibnamefont
  {Tkatchenko}}, \bibinfo {author} {\bibfnamefont {K.~R.}\ \bibnamefont
  {M\"{u}ller}}, \ and\ \bibinfo {author} {\bibfnamefont {R.~J.}\ \bibnamefont
  {Maurer}},\ }\href {\doibase 10.1038/s41467-019-12875-2} {\bibfield
  {journal} {\bibinfo  {journal} {Nature Communications}\ }\textbf {\bibinfo
  {volume} {10}},\ \bibinfo {pages} {5024} (\bibinfo {year}
  {2019})}\BibitemShut {NoStop}%
\bibitem [{\citenamefont {Zhai}\ \emph {et~al.}(2015)\citenamefont {Zhai},
  \citenamefont {Ha},\ and\ \citenamefont {Alexandrova}}]{SS:alexandrova}%
  \BibitemOpen
  \bibfield  {author} {\bibinfo {author} {\bibfnamefont {H.}~\bibnamefont
  {Zhai}}, \bibinfo {author} {\bibfnamefont {M.-A.}\ \bibnamefont {Ha}}, \ and\
  \bibinfo {author} {\bibfnamefont {A.~N.}\ \bibnamefont {Alexandrova}},\
  }\href {\doibase 10.1021/acs.jctc.5b00065} {\bibfield  {journal} {\bibinfo
  {journal} {J. Chem. Theory Comput.}\ }\textbf {\bibinfo {volume} {11}},\
  \bibinfo {pages} {2385} (\bibinfo {year} {2015})}\BibitemShut {NoStop}%
\bibitem [{\citenamefont {Todorovi{\'c}}\ \emph {et~al.}(2017)\citenamefont
  {Todorovi{\'c}}, \citenamefont {Gutmann}, \citenamefont {Corander},\ and\
  \citenamefont {Rinke}}]{active:rinke}%
  \BibitemOpen
  \bibfield  {author} {\bibinfo {author} {\bibfnamefont {M.}~\bibnamefont
  {Todorovi{\'c}}}, \bibinfo {author} {\bibfnamefont {M.}~\bibnamefont
  {Gutmann}}, \bibinfo {author} {\bibfnamefont {J.}~\bibnamefont {Corander}}, \
  and\ \bibinfo {author} {\bibfnamefont {P.}~\bibnamefont {Rinke}},\ }\href
  {\doibase 10.1038/s41524-019-0175-2} {\bibfield  {journal} {\bibinfo
  {journal} {npj Computational Materials}\ }\textbf {\bibinfo {volume} {5}},\
  \bibinfo {pages} {35} (\bibinfo {year} {2017})}\BibitemShut {NoStop}%
\bibitem [{\citenamefont {Yamashita}\ \emph {et~al.}(2018)\citenamefont
  {Yamashita}, \citenamefont {Sato}, \citenamefont {Kino}, \citenamefont
  {Miyake}, \citenamefont {Tsuda},\ and\ \citenamefont
  {Oguchi}}]{active:oguchi}%
  \BibitemOpen
  \bibfield  {author} {\bibinfo {author} {\bibfnamefont {T.}~\bibnamefont
  {Yamashita}}, \bibinfo {author} {\bibfnamefont {N.}~\bibnamefont {Sato}},
  \bibinfo {author} {\bibfnamefont {H.}~\bibnamefont {Kino}}, \bibinfo {author}
  {\bibfnamefont {T.}~\bibnamefont {Miyake}}, \bibinfo {author} {\bibfnamefont
  {K.}~\bibnamefont {Tsuda}}, \ and\ \bibinfo {author} {\bibfnamefont
  {T.}~\bibnamefont {Oguchi}},\ }\href {\doibase
  10.1103/PhysRevMaterials.2.013803} {\bibfield  {journal} {\bibinfo  {journal}
  {Phys. Rev. Materials}\ }\textbf {\bibinfo {volume} {2}},\ \bibinfo {pages}
  {013803} (\bibinfo {year} {2018})}\BibitemShut {NoStop}%
\bibitem [{\citenamefont {Schmitz}\ and\ \citenamefont
  {Christiansen}(2018)}]{active:ove}%
  \BibitemOpen
  \bibfield  {author} {\bibinfo {author} {\bibfnamefont {G.}~\bibnamefont
  {Schmitz}}\ and\ \bibinfo {author} {\bibfnamefont {O.}~\bibnamefont
  {Christiansen}},\ }\href {\doibase 10.1063/1.5009347} {\bibfield  {journal}
  {\bibinfo  {journal} {The Journal of Chemical Physics}\ }\textbf {\bibinfo
  {volume} {148}},\ \bibinfo {pages} {241704} (\bibinfo {year}
  {2018})}\BibitemShut {NoStop}%
\bibitem [{\citenamefont {Deringer}\ \emph
  {et~al.}(2018{\natexlab{a}})\citenamefont {Deringer}, \citenamefont
  {Pickard},\ and\ \citenamefont {Cs\'anyi}}]{GAPRSS:boron}%
  \BibitemOpen
  \bibfield  {author} {\bibinfo {author} {\bibfnamefont {V.~L.}\ \bibnamefont
  {Deringer}}, \bibinfo {author} {\bibfnamefont {C.~J.}\ \bibnamefont
  {Pickard}}, \ and\ \bibinfo {author} {\bibfnamefont {G.}~\bibnamefont
  {Cs\'anyi}},\ }\href {\doibase 10.1103/PhysRevLett.120.156001} {\bibfield
  {journal} {\bibinfo  {journal} {Phys. Rev. Lett.}\ }\textbf {\bibinfo
  {volume} {120}},\ \bibinfo {pages} {156001} (\bibinfo {year}
  {2018}{\natexlab{a}})}\BibitemShut {NoStop}%
\bibitem [{\citenamefont {Deringer}\ \emph
  {et~al.}(2018{\natexlab{b}})\citenamefont {Deringer}, \citenamefont
  {Proserpio}, \citenamefont {Cs\'anyi},\ and\ \citenamefont
  {Pickard}}]{GAPRSS:crystal}%
  \BibitemOpen
  \bibfield  {author} {\bibinfo {author} {\bibfnamefont {V.~L.}\ \bibnamefont
  {Deringer}}, \bibinfo {author} {\bibfnamefont {D.~M.}\ \bibnamefont
  {Proserpio}}, \bibinfo {author} {\bibfnamefont {G.}~\bibnamefont {Cs\'anyi}},
  \ and\ \bibinfo {author} {\bibfnamefont {C.~J.}\ \bibnamefont {Pickard}},\
  }\href {\doibase 10.1039/C8FD00034D} {\bibfield  {journal} {\bibinfo
  {journal} {Faraday Discuss.}\ }\textbf {\bibinfo {volume} {211}},\ \bibinfo
  {pages} {45} (\bibinfo {year} {2018}{\natexlab{b}})}\BibitemShut {NoStop}%
\bibitem [{\citenamefont {{Tong}}\ \emph {et~al.}(2018)\citenamefont {{Tong}},
  \citenamefont {{Xue}}, \citenamefont {{Lv}}, \citenamefont {{Wang}},\ and\
  \citenamefont {{Ma}}}]{activeSS:calypso}%
  \BibitemOpen
  \bibfield  {author} {\bibinfo {author} {\bibfnamefont {Q.}~\bibnamefont
  {{Tong}}}, \bibinfo {author} {\bibfnamefont {L.}~\bibnamefont {{Xue}}},
  \bibinfo {author} {\bibfnamefont {J.}~\bibnamefont {{Lv}}}, \bibinfo {author}
  {\bibfnamefont {Y.}~\bibnamefont {{Wang}}}, \ and\ \bibinfo {author}
  {\bibfnamefont {Y.}~\bibnamefont {{Ma}}},\ }\href {\doibase
  10.1039/C8FD00055G} {\bibfield  {journal} {\bibinfo  {journal} {Faraday
  Discussions}\ }\textbf {\bibinfo {volume} {211}},\ \bibinfo {pages} {31}
  (\bibinfo {year} {2018})}\BibitemShut {NoStop}%
\bibitem [{\citenamefont {Gubaev}\ \emph {et~al.}(2019)\citenamefont {Gubaev},
  \citenamefont {Podryabinkin}, \citenamefont {Hart},\ and\ \citenamefont
  {Shapeev}}]{activeSS:shapeev}%
  \BibitemOpen
  \bibfield  {author} {\bibinfo {author} {\bibfnamefont {K.}~\bibnamefont
  {Gubaev}}, \bibinfo {author} {\bibfnamefont {E.}~\bibnamefont
  {Podryabinkin}}, \bibinfo {author} {\bibfnamefont {G.}~\bibnamefont {Hart}},
  \ and\ \bibinfo {author} {\bibfnamefont {A.}~\bibnamefont {Shapeev}},\ }\href
  {\doibase 10.1016/j.commatsci.2018.09.031} {\bibfield  {journal} {\bibinfo
  {journal} {Comput. Mater. Sci.}\ }\textbf {\bibinfo {volume} {156}},\
  \bibinfo {pages} {148} (\bibinfo {year} {2019})}\BibitemShut {NoStop}%
\bibitem [{\citenamefont {Van~den Bossche}(2019)}]{SS_dftb:maxime}%
  \BibitemOpen
  \bibfield  {author} {\bibinfo {author} {\bibfnamefont {M.}~\bibnamefont
  {Van~den Bossche}},\ }\href {\doibase 10.1021/acs.jpca.9b00927} {\bibfield
  {journal} {\bibinfo  {journal} {J. Phys. Chem. A}\ }\textbf {\bibinfo
  {volume} {123}},\ \bibinfo {pages} {3038} (\bibinfo {year}
  {2019})}\BibitemShut {NoStop}%
\bibitem [{\citenamefont {Smith}\ \emph {et~al.}(2018)\citenamefont {Smith},
  \citenamefont {Nebgen}, \citenamefont {Lubbers}, \citenamefont {Isayev},\
  and\ \citenamefont {Roitberg}}]{activeFF:roitberg}%
  \BibitemOpen
  \bibfield  {author} {\bibinfo {author} {\bibfnamefont {J.~S.}\ \bibnamefont
  {Smith}}, \bibinfo {author} {\bibfnamefont {B.}~\bibnamefont {Nebgen}},
  \bibinfo {author} {\bibfnamefont {N.}~\bibnamefont {Lubbers}}, \bibinfo
  {author} {\bibfnamefont {O.}~\bibnamefont {Isayev}}, \ and\ \bibinfo {author}
  {\bibfnamefont {A.~E.}\ \bibnamefont {Roitberg}},\ }\href {\doibase
  10.1063/1.5023802} {\bibfield  {journal} {\bibinfo  {journal} {J. Chem.
  Phys.}\ }\textbf {\bibinfo {volume} {148}},\ \bibinfo {pages} {241733}
  (\bibinfo {year} {2018})}\BibitemShut {NoStop}%
\bibitem [{\citenamefont {Zhang}\ \emph {et~al.}(2019)\citenamefont {Zhang},
  \citenamefont {Lin}, \citenamefont {Wang}, \citenamefont {Car},\ and\
  \citenamefont {E}}]{activeFF:Zhang}%
  \BibitemOpen
  \bibfield  {author} {\bibinfo {author} {\bibfnamefont {L.}~\bibnamefont
  {Zhang}}, \bibinfo {author} {\bibfnamefont {D.-Y.}\ \bibnamefont {Lin}},
  \bibinfo {author} {\bibfnamefont {H.}~\bibnamefont {Wang}}, \bibinfo {author}
  {\bibfnamefont {R.}~\bibnamefont {Car}}, \ and\ \bibinfo {author}
  {\bibfnamefont {W.}~\bibnamefont {E}},\ }\href {\doibase
  10.1103/PhysRevMaterials.3.023804} {\bibfield  {journal} {\bibinfo  {journal}
  {Physical Review Materials}\ }\textbf {\bibinfo {volume} {3}},\ \bibinfo
  {pages} {023804} (\bibinfo {year} {2019})}\BibitemShut {NoStop}%
\bibitem [{\citenamefont {Kolsbjerg}\ \emph {et~al.}(2018)\citenamefont
  {Kolsbjerg}, \citenamefont {Peterson},\ and\ \citenamefont {Hammer}}]{LEA}%
  \BibitemOpen
  \bibfield  {author} {\bibinfo {author} {\bibfnamefont {E.~L.}\ \bibnamefont
  {Kolsbjerg}}, \bibinfo {author} {\bibfnamefont {A.~A.}\ \bibnamefont
  {Peterson}}, \ and\ \bibinfo {author} {\bibfnamefont {B.}~\bibnamefont
  {Hammer}},\ }\href {\doibase 10.1103/PhysRevB.97.195424} {\bibfield
  {journal} {\bibinfo  {journal} {Phys. Rev. B}\ }\textbf {\bibinfo {volume}
  {97}},\ \bibinfo {pages} {195424} (\bibinfo {year} {2018})}\BibitemShut
  {NoStop}%
\bibitem [{\citenamefont {Li}\ \emph {et~al.}(2015)\citenamefont {Li},
  \citenamefont {Kermode},\ and\ \citenamefont {De~Vita}}]{activeMD:Vita}%
  \BibitemOpen
  \bibfield  {author} {\bibinfo {author} {\bibfnamefont {Z.}~\bibnamefont
  {Li}}, \bibinfo {author} {\bibfnamefont {J.~R.}\ \bibnamefont {Kermode}}, \
  and\ \bibinfo {author} {\bibfnamefont {A.}~\bibnamefont {De~Vita}},\ }\href
  {\doibase 10.1103/PhysRevLett.114.096405} {\bibfield  {journal} {\bibinfo
  {journal} {Phys. Rev. Lett.}\ }\textbf {\bibinfo {volume} {114}},\ \bibinfo
  {pages} {096405} (\bibinfo {year} {2015})}\BibitemShut {NoStop}%
\bibitem [{\citenamefont {A.~Peterson}\ \emph {et~al.}(2017)\citenamefont
  {A.~Peterson}, \citenamefont {Christensen},\ and\ \citenamefont
  {Khorshidi}}]{activeMD:andrew}%
  \BibitemOpen
  \bibfield  {author} {\bibinfo {author} {\bibfnamefont {A.}~\bibnamefont
  {A.~Peterson}}, \bibinfo {author} {\bibfnamefont {R.}~\bibnamefont
  {Christensen}}, \ and\ \bibinfo {author} {\bibfnamefont {A.}~\bibnamefont
  {Khorshidi}},\ }\href {\doibase 10.1039/C7CP00375G} {\bibfield  {journal}
  {\bibinfo  {journal} {Phys. Chem. Chem. Phys.}\ }\textbf {\bibinfo {volume}
  {19}},\ \bibinfo {pages} {10978} (\bibinfo {year} {2017})}\BibitemShut
  {NoStop}%
\bibitem [{\citenamefont {Miwa}\ and\ \citenamefont
  {Ohno}(2017)}]{activeMD:Ohno}%
  \BibitemOpen
  \bibfield  {author} {\bibinfo {author} {\bibfnamefont {K.}~\bibnamefont
  {Miwa}}\ and\ \bibinfo {author} {\bibfnamefont {H.}~\bibnamefont {Ohno}},\
  }\href {\doibase 10.1103/PhysRevMaterials.1.053801} {\bibfield  {journal}
  {\bibinfo  {journal} {Phys. Rev. Materials}\ }\textbf {\bibinfo {volume}
  {1}},\ \bibinfo {pages} {053801} (\bibinfo {year} {2017})}\BibitemShut
  {NoStop}%
\bibitem [{\citenamefont {Podryabinkin}\ and\ \citenamefont
  {Shapeev}(2017)}]{activeMD:evgeny}%
  \BibitemOpen
  \bibfield  {author} {\bibinfo {author} {\bibfnamefont {E.~V.}\ \bibnamefont
  {Podryabinkin}}\ and\ \bibinfo {author} {\bibfnamefont {A.~V.}\ \bibnamefont
  {Shapeev}},\ }\href {\doibase
  https://doi.org/10.1016/j.commatsci.2017.08.031} {\bibfield  {journal}
  {\bibinfo  {journal} {Comput. Mater. Sci.}\ }\textbf {\bibinfo {volume}
  {140}},\ \bibinfo {pages} {171 } (\bibinfo {year} {2017})}\BibitemShut
  {NoStop}%
\bibitem [{\citenamefont {Jinnouchi}\ \emph {et~al.}(2019)\citenamefont
  {Jinnouchi}, \citenamefont {Karsai},\ and\ \citenamefont
  {Kresse}}]{activeMD:kresse}%
  \BibitemOpen
  \bibfield  {author} {\bibinfo {author} {\bibfnamefont {R.}~\bibnamefont
  {Jinnouchi}}, \bibinfo {author} {\bibfnamefont {F.}~\bibnamefont {Karsai}}, \
  and\ \bibinfo {author} {\bibfnamefont {G.}~\bibnamefont {Kresse}},\ }\href
  {\doibase 10.1103/PhysRevB.100.014105} {\bibfield  {journal} {\bibinfo
  {journal} {Phys. Rev. B}\ }\textbf {\bibinfo {volume} {100}},\ \bibinfo
  {pages} {014105} (\bibinfo {year} {2019})}\BibitemShut {NoStop}%
\bibitem [{\citenamefont {{Bernstein}}\ \emph {et~al.}(2019)\citenamefont
  {{Bernstein}}, \citenamefont {{Cs{\'a}nyi}},\ and\ \citenamefont
  {{Deringer}}}]{activeSS:deringer}%
  \BibitemOpen
  \bibfield  {author} {\bibinfo {author} {\bibfnamefont {N.}~\bibnamefont
  {{Bernstein}}}, \bibinfo {author} {\bibfnamefont {G.}~\bibnamefont
  {{Cs{\'a}nyi}}}, \ and\ \bibinfo {author} {\bibfnamefont {V.~L.}\
  \bibnamefont {{Deringer}}},\ }\href {\doibase 10.1038/s41524-019-0236-6}
  {\bibfield  {journal} {\bibinfo  {journal} {npj Computational Materials}\
  }\textbf {\bibinfo {volume} {5}},\ \bibinfo {pages} {99} (\bibinfo {year}
  {2019})}\BibitemShut {NoStop}%
\bibitem [{\citenamefont {Koistinen}\ \emph {et~al.}(2017)\citenamefont
  {Koistinen}, \citenamefont {Dagbjartsd\'{o}ttir}, \citenamefont
  {\'{A}sgeirsson}, \citenamefont {Vehtari},\ and\ \citenamefont
  {J\'{o}nsson}}]{NEB_hannes2017}%
  \BibitemOpen
  \bibfield  {author} {\bibinfo {author} {\bibfnamefont {O.-P.}\ \bibnamefont
  {Koistinen}}, \bibinfo {author} {\bibfnamefont {F.~B.}\ \bibnamefont
  {Dagbjartsd\'{o}ttir}}, \bibinfo {author} {\bibfnamefont {V.}~\bibnamefont
  {\'{A}sgeirsson}}, \bibinfo {author} {\bibfnamefont {A.}~\bibnamefont
  {Vehtari}}, \ and\ \bibinfo {author} {\bibfnamefont {H.}~\bibnamefont
  {J\'{o}nsson}},\ }\href {\doibase 10.1063/1.4986787} {\bibfield  {journal}
  {\bibinfo  {journal} {J. Chem. Phys.}\ }\textbf {\bibinfo {volume} {147}},\
  \bibinfo {pages} {152720} (\bibinfo {year} {2017})}\BibitemShut {NoStop}%
\bibitem [{\citenamefont {Garijo~del R\'{\i}o}\ \emph
  {et~al.}(2019)\citenamefont {Garijo~del R\'{\i}o}, \citenamefont
  {Mortensen},\ and\ \citenamefont {Jacobsen}}]{localOpt_karsten2019}%
  \BibitemOpen
  \bibfield  {author} {\bibinfo {author} {\bibfnamefont {E.}~\bibnamefont
  {Garijo~del R\'{\i}o}}, \bibinfo {author} {\bibfnamefont {J.~J.}\
  \bibnamefont {Mortensen}}, \ and\ \bibinfo {author} {\bibfnamefont {K.~W.}\
  \bibnamefont {Jacobsen}},\ }\href {\doibase 10.1103/PhysRevB.100.104103}
  {\bibfield  {journal} {\bibinfo  {journal} {Phys. Rev. B}\ }\textbf {\bibinfo
  {volume} {100}},\ \bibinfo {pages} {104103} (\bibinfo {year}
  {2019})}\BibitemShut {NoStop}%
\bibitem [{\citenamefont {J\o{}rgensen}\ \emph {et~al.}(2018)\citenamefont
  {J\o{}rgensen}, \citenamefont {Larsen}, \citenamefont {Jacobsen},\ and\
  \citenamefont {Hammer}}]{explo1}%
  \BibitemOpen
  \bibfield  {author} {\bibinfo {author} {\bibfnamefont {M.~S.}\ \bibnamefont
  {J\o{}rgensen}}, \bibinfo {author} {\bibfnamefont {U.~F.}\ \bibnamefont
  {Larsen}}, \bibinfo {author} {\bibfnamefont {K.~W.}\ \bibnamefont
  {Jacobsen}}, \ and\ \bibinfo {author} {\bibfnamefont {B.}~\bibnamefont
  {Hammer}},\ }\href {\doibase 10.1021/acs.jpca.8b00160} {\bibfield  {journal}
  {\bibinfo  {journal} {The Journal of Physical Chemistry A}\ }\textbf
  {\bibinfo {volume} {122}},\ \bibinfo {pages} {1504} (\bibinfo {year}
  {2018})}\BibitemShut {NoStop}%
\bibitem [{\citenamefont {Todorovi\'{c}}\ \emph {et~al.}(2019)\citenamefont
  {Todorovi\'{c}}, \citenamefont {Gutmann}, \citenamefont {Corander},\ and\
  \citenamefont {Rinke}}]{explo2}%
  \BibitemOpen
  \bibfield  {author} {\bibinfo {author} {\bibfnamefont {M.}~\bibnamefont
  {Todorovi\'{c}}}, \bibinfo {author} {\bibfnamefont {M.~U.}\ \bibnamefont
  {Gutmann}}, \bibinfo {author} {\bibfnamefont {J.}~\bibnamefont {Corander}}, \
  and\ \bibinfo {author} {\bibfnamefont {P.}~\bibnamefont {Rinke}},\ }\href
  {\doibase 10.1038/s41524-019-0175-2} {\bibfield  {journal} {\bibinfo
  {journal} {npj Computational Materials}\ }\textbf {\bibinfo {volume} {5}},\
  \bibinfo {pages} {35} (\bibinfo {year} {2019})}\BibitemShut {NoStop}%
\bibitem [{\citenamefont {J\o{}rgensen}\ \emph {et~al.}(2017)\citenamefont
  {J\o{}rgensen}, \citenamefont {Groves},\ and\ \citenamefont
  {Hammer}}]{cluster}%
  \BibitemOpen
  \bibfield  {author} {\bibinfo {author} {\bibfnamefont {M.~S.}\ \bibnamefont
  {J\o{}rgensen}}, \bibinfo {author} {\bibfnamefont {M.~N.}\ \bibnamefont
  {Groves}}, \ and\ \bibinfo {author} {\bibfnamefont {B.}~\bibnamefont
  {Hammer}},\ }\href {\doibase 10.1021/acs.jctc.6b01119} {\bibfield  {journal}
  {\bibinfo  {journal} {Journal of Chemical Theory and Computation}\ }\textbf
  {\bibinfo {volume} {13}},\ \bibinfo {pages} {1486} (\bibinfo {year}
  {2017})}\BibitemShut {NoStop}%
\bibitem [{\citenamefont {Jacobsen}\ \emph {et~al.}(2018)\citenamefont
  {Jacobsen}, \citenamefont {J\o{}rgensen},\ and\ \citenamefont
  {Hammer}}]{le1}%
  \BibitemOpen
  \bibfield  {author} {\bibinfo {author} {\bibfnamefont {T.~L.}\ \bibnamefont
  {Jacobsen}}, \bibinfo {author} {\bibfnamefont {M.~S.}\ \bibnamefont
  {J\o{}rgensen}}, \ and\ \bibinfo {author} {\bibfnamefont {B.}~\bibnamefont
  {Hammer}},\ }\href {\doibase 10.1103/PhysRevLett.120.026102} {\bibfield
  {journal} {\bibinfo  {journal} {Phys. Rev. Lett.}\ }\textbf {\bibinfo
  {volume} {120}},\ \bibinfo {pages} {026102} (\bibinfo {year}
  {2018})}\BibitemShut {NoStop}%
\bibitem [{\citenamefont {Chen}\ \emph {et~al.}(2018)\citenamefont {Chen},
  \citenamefont {J\o{}rgensen}, \citenamefont {Li},\ and\ \citenamefont
  {Hammer}}]{le2}%
  \BibitemOpen
  \bibfield  {author} {\bibinfo {author} {\bibfnamefont {X.}~\bibnamefont
  {Chen}}, \bibinfo {author} {\bibfnamefont {M.~S.}\ \bibnamefont
  {J\o{}rgensen}}, \bibinfo {author} {\bibfnamefont {J.}~\bibnamefont {Li}}, \
  and\ \bibinfo {author} {\bibfnamefont {B.}~\bibnamefont {Hammer}},\ }\href
  {\doibase 10.1021/acs.jctc.8b00149} {\bibfield  {journal} {\bibinfo
  {journal} {Journal of Chemical Theory and Computation}\ }\textbf {\bibinfo
  {volume} {14}},\ \bibinfo {pages} {3933} (\bibinfo {year}
  {2018})}\BibitemShut {NoStop}%
\bibitem [{\citenamefont {Meldgaard}\ \emph {et~al.}(2018)\citenamefont
  {Meldgaard}, \citenamefont {Kolsbjerg},\ and\ \citenamefont {Hammer}}]{le3}%
  \BibitemOpen
  \bibfield  {author} {\bibinfo {author} {\bibfnamefont {S.~A.}\ \bibnamefont
  {Meldgaard}}, \bibinfo {author} {\bibfnamefont {E.~L.}\ \bibnamefont
  {Kolsbjerg}}, \ and\ \bibinfo {author} {\bibfnamefont {B.}~\bibnamefont
  {Hammer}},\ }\href {\doibase 10.1063/1.5048290} {\bibfield  {journal}
  {\bibinfo  {journal} {The Journal of Chemical Physics}\ }\textbf {\bibinfo
  {volume} {149}},\ \bibinfo {pages} {134104} (\bibinfo {year}
  {2018})}\BibitemShut {NoStop}%
\bibitem [{\citenamefont {Chiriki}\ \emph {et~al.}(2019)\citenamefont
  {Chiriki}, \citenamefont {Christiansen},\ and\ \citenamefont
  {Hammer}}]{convex1}%
  \BibitemOpen
  \bibfield  {author} {\bibinfo {author} {\bibfnamefont {S.}~\bibnamefont
  {Chiriki}}, \bibinfo {author} {\bibfnamefont {M.-P.~V.}\ \bibnamefont
  {Christiansen}}, \ and\ \bibinfo {author} {\bibfnamefont {B.}~\bibnamefont
  {Hammer}},\ }\href {\doibase 10.1103/PhysRevB.100.235436} {\bibfield
  {journal} {\bibinfo  {journal} {Phys. Rev. B}\ }\textbf {\bibinfo {volume}
  {100}},\ \bibinfo {pages} {235436} (\bibinfo {year} {2019})}\BibitemShut
  {NoStop}%
\bibitem [{\citenamefont {S\o{}rensen}\ \emph {et~al.}(2018)\citenamefont
  {S\o{}rensen}, \citenamefont {J\o{}rgensen}, \citenamefont {Bruix},\ and\
  \citenamefont {Hammer}}]{convex2}%
  \BibitemOpen
  \bibfield  {author} {\bibinfo {author} {\bibfnamefont {K.~H.}\ \bibnamefont
  {S\o{}rensen}}, \bibinfo {author} {\bibfnamefont {M.~S.}\ \bibnamefont
  {J\o{}rgensen}}, \bibinfo {author} {\bibfnamefont {A.}~\bibnamefont {Bruix}},
  \ and\ \bibinfo {author} {\bibfnamefont {B.}~\bibnamefont {Hammer}},\ }\href
  {\doibase 10.1063/1.5023671} {\bibfield  {journal} {\bibinfo  {journal} {The
  Journal of Chemical Physics}\ }\textbf {\bibinfo {volume} {148}},\ \bibinfo
  {pages} {241734} (\bibinfo {year} {2018})}\BibitemShut {NoStop}%
\bibitem [{\citenamefont {J{\o}rgensen}\ \emph {et~al.}(2019)\citenamefont
  {J{\o}rgensen}, \citenamefont {Mortensen}, \citenamefont {Meldgaard},
  \citenamefont {Kolsbjerg}, \citenamefont {Jacobsen}, \citenamefont
  {S{\o}rensen},\ and\ \citenamefont {Hammer}}]{asla}%
  \BibitemOpen
  \bibfield  {author} {\bibinfo {author} {\bibfnamefont {M.~S.}\ \bibnamefont
  {J{\o}rgensen}}, \bibinfo {author} {\bibfnamefont {H.~L.}\ \bibnamefont
  {Mortensen}}, \bibinfo {author} {\bibfnamefont {S.~A.}\ \bibnamefont
  {Meldgaard}}, \bibinfo {author} {\bibfnamefont {E.~L.}\ \bibnamefont
  {Kolsbjerg}}, \bibinfo {author} {\bibfnamefont {T.~L.}\ \bibnamefont
  {Jacobsen}}, \bibinfo {author} {\bibfnamefont {K.~H.}\ \bibnamefont
  {S{\o}rensen}}, \ and\ \bibinfo {author} {\bibfnamefont {B.}~\bibnamefont
  {Hammer}},\ }\href {\doibase 10.1063/1.5108871} {\bibfield  {journal}
  {\bibinfo  {journal} {The Journal of Chemical Physics}\ }\textbf {\bibinfo
  {volume} {151}},\ \bibinfo {pages} {054111} (\bibinfo {year}
  {2019})}\BibitemShut {NoStop}%
\bibitem [{\citenamefont {Schnadt}\ \emph {et~al.}(2009)\citenamefont
  {Schnadt}, \citenamefont {Knudsen}, \citenamefont {Hu}, \citenamefont
  {Michaelides}, \citenamefont {Vang}, \citenamefont {Reuter}, \citenamefont
  {Li}, \citenamefont {L\ae{}gsgaard}, \citenamefont {Scheffler},\ and\
  \citenamefont {Besenbacher}}]{schnadt_prb_2009}%
  \BibitemOpen
  \bibfield  {author} {\bibinfo {author} {\bibfnamefont {J.}~\bibnamefont
  {Schnadt}}, \bibinfo {author} {\bibfnamefont {J.}~\bibnamefont {Knudsen}},
  \bibinfo {author} {\bibfnamefont {X.~L.}\ \bibnamefont {Hu}}, \bibinfo
  {author} {\bibfnamefont {A.}~\bibnamefont {Michaelides}}, \bibinfo {author}
  {\bibfnamefont {R.~T.}\ \bibnamefont {Vang}}, \bibinfo {author}
  {\bibfnamefont {K.}~\bibnamefont {Reuter}}, \bibinfo {author} {\bibfnamefont
  {Z.}~\bibnamefont {Li}}, \bibinfo {author} {\bibfnamefont {E.}~\bibnamefont
  {L\ae{}gsgaard}}, \bibinfo {author} {\bibfnamefont {M.}~\bibnamefont
  {Scheffler}}, \ and\ \bibinfo {author} {\bibfnamefont {F.}~\bibnamefont
  {Besenbacher}},\ }\href {\doibase 10.1103/PhysRevB.80.075424} {\bibfield
  {journal} {\bibinfo  {journal} {Phys. Rev. B}\ }\textbf {\bibinfo {volume}
  {80}},\ \bibinfo {pages} {075424} (\bibinfo {year} {2009})}\BibitemShut
  {NoStop}%
\bibitem [{\citenamefont {Martin}\ \emph {et~al.}(2014)\citenamefont {Martin},
  \citenamefont {Klacar}, \citenamefont {Gr\"{o}nbeck}, \citenamefont
  {Knudsen}, \citenamefont {Schnadt}, \citenamefont {Blomberg}, \citenamefont
  {Gustafson},\ and\ \citenamefont {Lundgren}}]{martin_j_phys_chem_c_2014}%
  \BibitemOpen
  \bibfield  {author} {\bibinfo {author} {\bibfnamefont {N.~M.}\ \bibnamefont
  {Martin}}, \bibinfo {author} {\bibfnamefont {S.}~\bibnamefont {Klacar}},
  \bibinfo {author} {\bibfnamefont {H.}~\bibnamefont {Gr\"{o}nbeck}}, \bibinfo
  {author} {\bibfnamefont {J.}~\bibnamefont {Knudsen}}, \bibinfo {author}
  {\bibfnamefont {J.}~\bibnamefont {Schnadt}}, \bibinfo {author} {\bibfnamefont
  {S.}~\bibnamefont {Blomberg}}, \bibinfo {author} {\bibfnamefont
  {J.}~\bibnamefont {Gustafson}}, \ and\ \bibinfo {author} {\bibfnamefont
  {E.}~\bibnamefont {Lundgren}},\ }\href {\doibase 10.1021/jp504387p}
  {\bibfield  {journal} {\bibinfo  {journal} {The Journal of Physical Chemistry
  C}\ }\textbf {\bibinfo {volume} {118}},\ \bibinfo {pages} {15324} (\bibinfo
  {year} {2014})}\BibitemShut {NoStop}%
\bibitem [{\citenamefont {Michaelides}\ \emph {et~al.}(2005)\citenamefont
  {Michaelides}, \citenamefont {Reuter},\ and\ \citenamefont
  {Scheffler}}]{ag111_reuter}%
  \BibitemOpen
  \bibfield  {author} {\bibinfo {author} {\bibfnamefont {A.}~\bibnamefont
  {Michaelides}}, \bibinfo {author} {\bibfnamefont {K.}~\bibnamefont {Reuter}},
  \ and\ \bibinfo {author} {\bibfnamefont {M.}~\bibnamefont {Scheffler}},\
  }\href {\doibase 10.1116/1.2049302} {\bibfield  {journal} {\bibinfo
  {journal} {Journal of Vacuum Science \& Technology A}\ }\textbf {\bibinfo
  {volume} {23}},\ \bibinfo {pages} {1487} (\bibinfo {year} {2005})},\ \Eprint
  {http://arxiv.org/abs/https://doi.org/10.1116/1.2049302}
  {https://doi.org/10.1116/1.2049302} \BibitemShut {NoStop}%
\bibitem [{\citenamefont {Meldgaard}\ \emph {et~al.}(2020)\citenamefont
  {Meldgaard}, \citenamefont {Mortensen}, \citenamefont {J{\o}rgensen},\ and\
  \citenamefont {Hammer}}]{asla_pseudo_3D}%
  \BibitemOpen
  \bibfield  {author} {\bibinfo {author} {\bibfnamefont {S.~A.}\ \bibnamefont
  {Meldgaard}}, \bibinfo {author} {\bibfnamefont {H.~L.}\ \bibnamefont
  {Mortensen}}, \bibinfo {author} {\bibfnamefont {M.~S.}\ \bibnamefont
  {J{\o}rgensen}}, \ and\ \bibinfo {author} {\bibfnamefont {B.}~\bibnamefont
  {Hammer}},\ }\href {\doibase 10.1088/1361-648x/ab94f2} {\bibfield  {journal}
  {\bibinfo  {journal} {Journal of Physics: Condensed Matter}\ }\textbf
  {\bibinfo {volume} {32}},\ \bibinfo {pages} {404005} (\bibinfo {year}
  {2020})}\BibitemShut {NoStop}%
\bibitem [{\citenamefont {Christiansen}\ \emph {et~al.}()\citenamefont
  {Christiansen}, \citenamefont {Mortensen}, \citenamefont {Meldgaard},\ and\
  \citenamefont {Hammer}}]{asla_representation}%
  \BibitemOpen
  \bibfield  {author} {\bibinfo {author} {\bibfnamefont {M.-P.~V.}\
  \bibnamefont {Christiansen}}, \bibinfo {author} {\bibfnamefont {H.~L.}\
  \bibnamefont {Mortensen}}, \bibinfo {author} {\bibfnamefont {S.~A.}\
  \bibnamefont {Meldgaard}}, \ and\ \bibinfo {author} {\bibfnamefont
  {B.}~\bibnamefont {Hammer}},\ }\href@noop {} {\ }\bibinfo {note} {Submitted
  to J. Chem. Phys.}\BibitemShut {Stop}%
\bibitem [{\citenamefont {Mortensen}\ \emph {et~al.}(2005)\citenamefont
  {Mortensen}, \citenamefont {Hansen},\ and\ \citenamefont {Jacobsen}}]{gpaw1}%
  \BibitemOpen
  \bibfield  {author} {\bibinfo {author} {\bibfnamefont {J.~J.}\ \bibnamefont
  {Mortensen}}, \bibinfo {author} {\bibfnamefont {L.~B.}\ \bibnamefont
  {Hansen}}, \ and\ \bibinfo {author} {\bibfnamefont {K.~W.}\ \bibnamefont
  {Jacobsen}},\ }\href {\doibase 10.1103/PhysRevB.71.035109} {\bibfield
  {journal} {\bibinfo  {journal} {Phys. Rev. B}\ }\textbf {\bibinfo {volume}
  {71}},\ \bibinfo {pages} {035109} (\bibinfo {year} {2005})}\BibitemShut
  {NoStop}%
\bibitem [{\citenamefont {Enkovaara}\ \emph {et~al.}(2010)\citenamefont
  {Enkovaara}, \citenamefont {Rostgaard}, \citenamefont {Mortensen},
  \citenamefont {Chen}, \citenamefont {Du{\l}ak}, \citenamefont {Ferrighi},
  \citenamefont {Gavnholt}, \citenamefont {Glinsvad}, \citenamefont {Haikola},
  \citenamefont {Hansen}, \citenamefont {Kristoffersen}, \citenamefont
  {Kuisma}, \citenamefont {Larsen}, \citenamefont {Lehtovaara}, \citenamefont
  {Ljungberg}, \citenamefont {Lopez-Acevedo}, \citenamefont {Moses},
  \citenamefont {Ojanen}, \citenamefont {Olsen}, \citenamefont {Petzold},
  \citenamefont {Romero}, \citenamefont {Stausholm-M{\o}ller}, \citenamefont
  {Strange}, \citenamefont {Tritsaris}, \citenamefont {Vanin}, \citenamefont
  {Walter}, \citenamefont {Hammer}, \citenamefont {H\"{a}kkinen}, \citenamefont
  {Madsen}, \citenamefont {Nieminen}, \citenamefont {N{\o}rskov}, \citenamefont
  {Puska}, \citenamefont {Rantala}, \citenamefont {Schi{\o}tz}, \citenamefont
  {Thygesen},\ and\ \citenamefont {Jacobsen}}]{gpaw2}%
  \BibitemOpen
  \bibfield  {author} {\bibinfo {author} {\bibfnamefont {J.}~\bibnamefont
  {Enkovaara}}, \bibinfo {author} {\bibfnamefont {C.}~\bibnamefont
  {Rostgaard}}, \bibinfo {author} {\bibfnamefont {J.~J.}\ \bibnamefont
  {Mortensen}}, \bibinfo {author} {\bibfnamefont {J.}~\bibnamefont {Chen}},
  \bibinfo {author} {\bibfnamefont {M.}~\bibnamefont {Du{\l}ak}}, \bibinfo
  {author} {\bibfnamefont {L.}~\bibnamefont {Ferrighi}}, \bibinfo {author}
  {\bibfnamefont {J.}~\bibnamefont {Gavnholt}}, \bibinfo {author}
  {\bibfnamefont {C.}~\bibnamefont {Glinsvad}}, \bibinfo {author}
  {\bibfnamefont {V.}~\bibnamefont {Haikola}}, \bibinfo {author} {\bibfnamefont
  {H.~A.}\ \bibnamefont {Hansen}}, \bibinfo {author} {\bibfnamefont {H.~H.}\
  \bibnamefont {Kristoffersen}}, \bibinfo {author} {\bibfnamefont
  {M.}~\bibnamefont {Kuisma}}, \bibinfo {author} {\bibfnamefont {A.~H.}\
  \bibnamefont {Larsen}}, \bibinfo {author} {\bibfnamefont {L.}~\bibnamefont
  {Lehtovaara}}, \bibinfo {author} {\bibfnamefont {M.}~\bibnamefont
  {Ljungberg}}, \bibinfo {author} {\bibfnamefont {O.}~\bibnamefont
  {Lopez-Acevedo}}, \bibinfo {author} {\bibfnamefont {P.~G.}\ \bibnamefont
  {Moses}}, \bibinfo {author} {\bibfnamefont {J.}~\bibnamefont {Ojanen}},
  \bibinfo {author} {\bibfnamefont {T.}~\bibnamefont {Olsen}}, \bibinfo
  {author} {\bibfnamefont {V.}~\bibnamefont {Petzold}}, \bibinfo {author}
  {\bibfnamefont {N.~A.}\ \bibnamefont {Romero}}, \bibinfo {author}
  {\bibfnamefont {J.}~\bibnamefont {Stausholm-M{\o}ller}}, \bibinfo {author}
  {\bibfnamefont {M.}~\bibnamefont {Strange}}, \bibinfo {author} {\bibfnamefont
  {G.~A.}\ \bibnamefont {Tritsaris}}, \bibinfo {author} {\bibfnamefont
  {M.}~\bibnamefont {Vanin}}, \bibinfo {author} {\bibfnamefont
  {M.}~\bibnamefont {Walter}}, \bibinfo {author} {\bibfnamefont
  {B.}~\bibnamefont {Hammer}}, \bibinfo {author} {\bibfnamefont
  {H.}~\bibnamefont {H\"{a}kkinen}}, \bibinfo {author} {\bibfnamefont
  {G.~K.~H.}\ \bibnamefont {Madsen}}, \bibinfo {author} {\bibfnamefont {R.~M.}\
  \bibnamefont {Nieminen}}, \bibinfo {author} {\bibfnamefont {J.~K.}\
  \bibnamefont {N{\o}rskov}}, \bibinfo {author} {\bibfnamefont
  {M.}~\bibnamefont {Puska}}, \bibinfo {author} {\bibfnamefont {T.~T.}\
  \bibnamefont {Rantala}}, \bibinfo {author} {\bibfnamefont {J.}~\bibnamefont
  {Schi{\o}tz}}, \bibinfo {author} {\bibfnamefont {K.~S.}\ \bibnamefont
  {Thygesen}}, \ and\ \bibinfo {author} {\bibfnamefont {K.~W.}\ \bibnamefont
  {Jacobsen}},\ }\href {\doibase 10.1088/0953-8984/22/25/253202} {\bibfield
  {journal} {\bibinfo  {journal} {Journal of Physics: Condensed Matter}\
  }\textbf {\bibinfo {volume} {22}},\ \bibinfo {pages} {253202} (\bibinfo
  {year} {2010})}\BibitemShut {NoStop}%
\bibitem [{\citenamefont {Rasmussen}\ and\ \citenamefont
  {Williams}(2005{\natexlab{a}})}]{GPs_for_ML}%
  \BibitemOpen
  \bibfield  {author} {\bibinfo {author} {\bibfnamefont {C.~E.}\ \bibnamefont
  {Rasmussen}}\ and\ \bibinfo {author} {\bibfnamefont {C.~K.~I.}\ \bibnamefont
  {Williams}},\ }\href@noop {} {\emph {\bibinfo {title} {Gaussian Processes for
  Machine Learning}}}\ (\bibinfo  {publisher} {The MIT Press},\ \bibinfo {year}
  {2005})\BibitemShut {NoStop}%
\bibitem [{\citenamefont {Oganov}\ and\ \citenamefont
  {Valle}(2009)}]{oganov_valle_2009}%
  \BibitemOpen
  \bibfield  {author} {\bibinfo {author} {\bibfnamefont {A.~R.}\ \bibnamefont
  {Oganov}}\ and\ \bibinfo {author} {\bibfnamefont {M.}~\bibnamefont {Valle}},\
  }\href {\doibase 10.1063/1.3079326} {\bibfield  {journal} {\bibinfo
  {journal} {The Journal of Chemical Physics}\ }\textbf {\bibinfo {volume}
  {130}},\ \bibinfo {pages} {104504} (\bibinfo {year} {2009})}\BibitemShut
  {NoStop}%
\bibitem [{\citenamefont {Bisbo}\ and\ \citenamefont
  {Hammer}(2020)}]{bisbo_hammer_2020}%
  \BibitemOpen
  \bibfield  {author} {\bibinfo {author} {\bibfnamefont {M.~K.}\ \bibnamefont
  {Bisbo}}\ and\ \bibinfo {author} {\bibfnamefont {B.}~\bibnamefont {Hammer}},\
  }\href {\doibase 10.1103/PhysRevLett.124.086102} {\bibfield  {journal}
  {\bibinfo  {journal} {Phys. Rev. Lett.}\ }\textbf {\bibinfo {volume} {124}},\
  \bibinfo {pages} {086102} (\bibinfo {year} {2020})}\BibitemShut {NoStop}%
\bibitem [{\citenamefont {Rasmussen}\ and\ \citenamefont
  {Williams}(2005{\natexlab{b}})}]{GP:Rasmussen}%
  \BibitemOpen
  \bibfield  {author} {\bibinfo {author} {\bibfnamefont {C.~E.}\ \bibnamefont
  {Rasmussen}}\ and\ \bibinfo {author} {\bibfnamefont {C.~K.~I.}\ \bibnamefont
  {Williams}},\ }\href@noop {} {\emph {\bibinfo {title} {Gaussian Processes for
  Machine Learning}}}\ (\bibinfo  {publisher} {The MIT Press},\ \bibinfo {year}
  {2005})\BibitemShut {NoStop}%
\bibitem [{\citenamefont {Aradi}\ \emph {et~al.}(2007)\citenamefont {Aradi},
  \citenamefont {Hourahine},\ and\ \citenamefont {Frauenheim}}]{dftb_plus}%
  \BibitemOpen
  \bibfield  {author} {\bibinfo {author} {\bibfnamefont {B.}~\bibnamefont
  {Aradi}}, \bibinfo {author} {\bibfnamefont {B.}~\bibnamefont {Hourahine}}, \
  and\ \bibinfo {author} {\bibfnamefont {T.}~\bibnamefont {Frauenheim}},\
  }\href {\doibase 10.1021/jp070186p} {\bibfield  {journal} {\bibinfo
  {journal} {The Journal of Physical Chemistry A}\ }\textbf {\bibinfo {volume}
  {111}},\ \bibinfo {pages} {5678} (\bibinfo {year} {2007})}\BibitemShut
  {NoStop}%
\bibitem [{\citenamefont {Larsen}\ \emph {et~al.}(2017)\citenamefont {Larsen},
  \citenamefont {Mortensen}, \citenamefont {Blomqvist}, \citenamefont
  {Castelli}, \citenamefont {Christensen}, \citenamefont {Dułak},
  \citenamefont {Friis}, \citenamefont {Groves}, \citenamefont {Hammer},
  \citenamefont {Hargus}, \citenamefont {Hermes}, \citenamefont {Jennings},
  \citenamefont {Jensen}, \citenamefont {Kermode}, \citenamefont {Kitchin},
  \citenamefont {Kolsbjerg}, \citenamefont {Kubal}, \citenamefont {Kaasbjerg},
  \citenamefont {Lysgaard}, \citenamefont {Maronsson}, \citenamefont {Maxson},
  \citenamefont {Olsen}, \citenamefont {Pastewka}, \citenamefont {Peterson},
  \citenamefont {Rostgaard}, \citenamefont {Schi{\o}tz}, \citenamefont
  {Sch{\"u}tt}, \citenamefont {Strange}, \citenamefont {Thygesen},
  \citenamefont {Vegge}, \citenamefont {Vilhelmsen}, \citenamefont {Walter},
  \citenamefont {Zeng},\ and\ \citenamefont {Jacobsen}}]{ase}%
  \BibitemOpen
  \bibfield  {author} {\bibinfo {author} {\bibfnamefont {A.~H.}\ \bibnamefont
  {Larsen}}, \bibinfo {author} {\bibfnamefont {J.~J.}\ \bibnamefont
  {Mortensen}}, \bibinfo {author} {\bibfnamefont {J.}~\bibnamefont
  {Blomqvist}}, \bibinfo {author} {\bibfnamefont {I.~E.}\ \bibnamefont
  {Castelli}}, \bibinfo {author} {\bibfnamefont {R.}~\bibnamefont
  {Christensen}}, \bibinfo {author} {\bibfnamefont {M.}~\bibnamefont {Dułak}},
  \bibinfo {author} {\bibfnamefont {J.}~\bibnamefont {Friis}}, \bibinfo
  {author} {\bibfnamefont {M.~N.}\ \bibnamefont {Groves}}, \bibinfo {author}
  {\bibfnamefont {B.}~\bibnamefont {Hammer}}, \bibinfo {author} {\bibfnamefont
  {C.}~\bibnamefont {Hargus}}, \bibinfo {author} {\bibfnamefont {E.~D.}\
  \bibnamefont {Hermes}}, \bibinfo {author} {\bibfnamefont {P.~C.}\
  \bibnamefont {Jennings}}, \bibinfo {author} {\bibfnamefont {P.~B.}\
  \bibnamefont {Jensen}}, \bibinfo {author} {\bibfnamefont {J.}~\bibnamefont
  {Kermode}}, \bibinfo {author} {\bibfnamefont {J.~R.}\ \bibnamefont
  {Kitchin}}, \bibinfo {author} {\bibfnamefont {E.~L.}\ \bibnamefont
  {Kolsbjerg}}, \bibinfo {author} {\bibfnamefont {J.}~\bibnamefont {Kubal}},
  \bibinfo {author} {\bibfnamefont {K.}~\bibnamefont {Kaasbjerg}}, \bibinfo
  {author} {\bibfnamefont {S.}~\bibnamefont {Lysgaard}}, \bibinfo {author}
  {\bibfnamefont {J.~B.}\ \bibnamefont {Maronsson}}, \bibinfo {author}
  {\bibfnamefont {T.}~\bibnamefont {Maxson}}, \bibinfo {author} {\bibfnamefont
  {T.}~\bibnamefont {Olsen}}, \bibinfo {author} {\bibfnamefont
  {L.}~\bibnamefont {Pastewka}}, \bibinfo {author} {\bibfnamefont
  {A.}~\bibnamefont {Peterson}}, \bibinfo {author} {\bibfnamefont
  {C.}~\bibnamefont {Rostgaard}}, \bibinfo {author} {\bibfnamefont
  {J.}~\bibnamefont {Schi{\o}tz}}, \bibinfo {author} {\bibfnamefont
  {O.}~\bibnamefont {Sch{\"u}tt}}, \bibinfo {author} {\bibfnamefont
  {M.}~\bibnamefont {Strange}}, \bibinfo {author} {\bibfnamefont {K.~S.}\
  \bibnamefont {Thygesen}}, \bibinfo {author} {\bibfnamefont {T.}~\bibnamefont
  {Vegge}}, \bibinfo {author} {\bibfnamefont {L.}~\bibnamefont {Vilhelmsen}},
  \bibinfo {author} {\bibfnamefont {M.}~\bibnamefont {Walter}}, \bibinfo
  {author} {\bibfnamefont {Z.}~\bibnamefont {Zeng}}, \ and\ \bibinfo {author}
  {\bibfnamefont {K.~W.}\ \bibnamefont {Jacobsen}},\ }\href
  {http://stacks.iop.org/0953-8984/29/i=27/a=273002} {\bibfield  {journal}
  {\bibinfo  {journal} {J. Phys. Condens. Matter}\ }\textbf {\bibinfo {volume}
  {29}},\ \bibinfo {pages} {273002} (\bibinfo {year} {2017})}\BibitemShut
  {NoStop}%
\bibitem [{\citenamefont {Perdew}\ \emph {et~al.}(1996)\citenamefont {Perdew},
  \citenamefont {Burke},\ and\ \citenamefont {Ernzerhof}}]{pbe}%
  \BibitemOpen
  \bibfield  {author} {\bibinfo {author} {\bibfnamefont {J.~P.}\ \bibnamefont
  {Perdew}}, \bibinfo {author} {\bibfnamefont {K.}~\bibnamefont {Burke}}, \
  and\ \bibinfo {author} {\bibfnamefont {M.}~\bibnamefont {Ernzerhof}},\ }\href
  {\doibase 10.1103/PhysRevLett.77.3865} {\bibfield  {journal} {\bibinfo
  {journal} {Phys. Rev. Lett.}\ }\textbf {\bibinfo {volume} {77}},\ \bibinfo
  {pages} {3865} (\bibinfo {year} {1996})}\BibitemShut {NoStop}%
\bibitem [{\citenamefont {Reuter}\ and\ \citenamefont
  {Scheffler}(2003)}]{reuter_prb_2003}%
  \BibitemOpen
  \bibfield  {author} {\bibinfo {author} {\bibfnamefont {K.}~\bibnamefont
  {Reuter}}\ and\ \bibinfo {author} {\bibfnamefont {M.}~\bibnamefont
  {Scheffler}},\ }\href {\doibase 10.1103/PhysRevB.68.045407} {\bibfield
  {journal} {\bibinfo  {journal} {Phys. Rev. B}\ }\textbf {\bibinfo {volume}
  {68}},\ \bibinfo {pages} {045407} (\bibinfo {year} {2003})}\BibitemShut
  {NoStop}%
\end{thebibliography}%

\end{document}